\documentclass[vecphys]{svmult}

\usepackage{makeidx}         
\usepackage{graphicx}        
\usepackage{multicol}        
\usepackage{cite}            
\usepackage[bottom]{footmisc}

\makeindex             

\begin{document}

\title{Dynamics of fully coupled rotators with 
unimodal and bimodal frequency distribution}
\titlerunning{Dynamics of fully coupled rotators}
\author{Simona Olmi \and Alessandro Torcini}
\institute{CNR - Consiglio Nazionale delle Ricerche - Istituto dei Sistemi
Complessi, via Madonna del Piano 10, I-50019 Sesto Fiorentino, Italy
\and INFN Sez. Firenze, via Sansone, 1 - I-50019 Sesto Fiorentino, Italy
\texttt{simona.olmi@fi.isc.cnr.it}
\texttt{alessandro.torcini@cnr.it}
}

\maketitle

\begin{abstract}
We analyze the synchronization transition of a  globally coupled 
network of $N$ phase oscillators with inertia (rotators) whose natural frequencies
 are unimodally or bimodally distributed.  In the unimodal case, the system exhibits a 
discontinuous hysteretic transition from an incoherent to a partially synchronized (PS) state. 
For sufficiently large inertia, the system reveals the coexistence of a PS state and of a 
standing wave (SW) solution. In the bimodal case, the hysteretic 
synchronization transition involves several states. Namely, the system 
becomes coherent passing through traveling waves (TWs), SWs
and finally arriving to a PS regime. The transition to the PS state from the SW occurs always at 
the same coupling, independently of the system size, while its value increases linearly with the inertia. 
On the other hand the critical coupling required to observe TWs  and SWs increases with $N$ suggesting 
that  in the thermodynamic limit the transition from incoherence to PS will occur without any intermediate 
states. Finally a linear stability analysis reveals that the system is hysteretic not only at the level 
of macroscopic indicators, but also microscopically as verified by measuring the maximal Lyapunov exponent.\end{abstract}

\noindent

\section{Introduction}
\index{Kuramoto model with inertia} \index{phase oscillator}

The renowned Kuramoto model~\cite{kura} for phase oscillators was generalized in 
1997 by Tanaka, Lichtenberg and Oishi (TLO)\cite{tanaka1997first,tanaka1997self} 
by including an additional inertial term. The TLO model revealed, at variance with the usual Kuramoto model, 
first order synchronization transitions even for unimodal distributions of the natural frequencies. 
TLO have been inspired in their extension by a work of Ermentrout published in 1991~\cite{ermentrout1991}; 
in this paper Ermentrout has introduced a pulse coupled phase oscillator model with inertia to mimic the 
{\it perfect synchrony} achieved by a specific type of fireflies, the {\it Pteroptix Malaccae} 
(but also by certain species of crickets and humans). 
The peculiarity of these fireflies is that they are able to synchronize their flashing activity to some 
forcing frequency (even quite distinct from  their own intrinsic flashing frequency) with an almost zero phase lag. 
This happens because they adapt their period of oscillation to that of the driving oscillator. 
After his introduction, the Kuramoto model with inertia has been employed to describe synchronization phenomena 
in crowd synchrony on London’s Millennium bridge~\cite{strogatz2005}, as well as in Huygen’s pendulum clocks 
\cite{bennett2002}. Furthermore, phase oscillators with inertia (rotators) have recently found application 
in the study the self-synchronization in power and smart grids
~\cite{salam1984,filatrella2008,rohden2012,dorfler2013,nishikawa2015}, as well as in the analysis of 
disordered arrays of underdamped Josephson junctions~\cite{trees2005}.
Cluster explosive synchronization has been reported for an adaptive network of
Kuramoto oscillators with inertia, where the natural frequency of each oscillator 
is assumed to be proportional to the degree of the corresponding node~\cite{Ji2013}.
Rotators arranged in two symmetrically coupled populations
have recently revealed the emergence of intermittent chaotic chimeras~\cite{olmi2015},
imperfect chimera states have been found in a ring with nonlocal coupling~\cite{jaros2015}, and transient waves 
have been observed in regular lattices~\cite{jorg2015}.

There is a wide literature devoted to coupled rotators~\index{rotators} with an unimodal
frequency distribution, however only a really limited number of
studies have been devoted to this model with a bimodal distribution,
despite the subject being extremely relevant for the modelization of the power grids \index{power grids}~\cite{filatrella2008,rohden2012,olmi2014}. To our knowledge 
the synchronization transition \index{synchronization transition} in populations of globally coupled rotators with bimodal 
distribution\index{bimodal frequency distribution} \index{inertia} has been previously analyzed only by Acebr\'{o}n et al. in~\cite{acebron2000}. 
More specifically, the authors considered a model
with white noise and a distribution composed by two $\delta$-functions localized
at $\pm \Omega_0$. As suggested in~\cite{martens2009}, the presence of noise 
blurs the $\delta$-functions in bell-shaped functions analogous
to Gaussian distributions. Therefore one expects a similar phenomenology
to the one observable for deterministic systems with bimodal Gaussian distributions
of the frequencies. A multiscale analysis of the model, in the limit of sufficiently large $\Omega_0$, reveals the emergence from the incoherent state
of stable standing wave solutions (SWs)~\index{standing waves} and of unstable traveling wave solutions (TWs)~\index{traveling waves} via supercritical 
bifurcations, while partially synchronized stationary states (PSs)~\index{partial synchronization} 
bifurcates subcritically from incoherence. 
However, the authors affirm that in the considered limit the bifurcation diagram coincides with that of 
the usual Kuramoto model without inertia~ \cite{acebron1998}.
 
In this article we analyze the synchronization transitions observable
for unimodal and bimodal frequency distributions for a population of globally coupled
rotators in a fully deterministic system. In particular, we will analyze the 
influence of inertia and of finite size effects on the synchronization
transitions. Moreover, we will study the macroscopic and microscopic characteristics of the different regimes 
emerging during adiabatic increase and decrease of the coupling among the rotators. In particular, Section 1.2 
will be devoted to the introduction of the model, of the indicators used to characterize the synchronization 
transition, and of the different protocols employed to perform adiabatic simulations. The Lyapunov linear stability 
analysis is introduced in SubSection 1.2.1. The results for unimodal distributions are reported in Sect. 1.3, 
with a particular emphasis to the TLO mean field
theory and its extension to any generic state observable within the hysteretic region 
(SubSect. 1.3.2). The emergence of clusters of locked and whirling oscillators is described in details in 
SubSect. 1.3.3. The dynamics of the network for bimodal distributions is analyzed in Section 1.4.
In particular, SubSect. 1.4.1 is devoted
to two non overlapping  distributions and SubSect. 1.4.2 to largely overlapping Gaussian distributions. 
Sect. 1.5 report the result of linear stability analysis for 
the considered distributions. Finally, in Sect. 1.6 the reported results are briefly summarized and discussed.

\section{Model and Indicators}
\index{Kuramoto model with inertia} \index{synchronization order parameter} \index{synchronization}

By following Refs.~\cite{tanaka1997self,tanaka1997first},
we study the following version of the Kuramoto model with inertia for $N$ fully coupled
rotators~\index{Kuramoto model with inertia}:
\begin{equation}
\label{eqPRL}
 m\ddot{\theta}_i + \dot{\theta}_i=\Omega_i + \frac{K}{N}\sum_j \sin(\theta_j-\theta_i)
\end{equation}
where $\theta_i$ and $\Omega_i$ are, respectively, the instantaneous phase and
the natural frequency of the $i$-th oscillator, $K$ is the coupling.
In the following we will consider random natural frequencies $\Omega_i$
Gaussian distributed according to:
an unimodal distribution $g(\Omega)=\frac{1}{\sqrt{2\pi}} {\rm e}^{-\frac{\Omega^2}{2}}$ 
with zero average and an unitary standard deviation
or a bimodal symmetric distribution $g(\Omega)=\frac{1}{2\sqrt{2\pi}} 
\left[ {\rm e}^{-\frac{(\Omega-\Omega_0)^2}{2}} +  {\rm e}^{-\frac{(\Omega+\Omega_0)^2}{2}}\right] $,
which is the overlap of two Gaussians with unitary standard deviation and
with the peaks located at a distance $2 \Omega_0$. 

To measure the level of coherence between the oscillators, we
employ the complex order parameter ~\cite{winfree}
\begin{equation}
\label{order_parameter}
 r(t)e^{i\phi(t)}=\frac{1}{N}\sum_j e^{i\theta_j} \enskip ;
\end{equation}
where $r(t) \in [0:1]$ is the modulus and $\phi(t)$ the phase of the 
macroscopic indicator. An asynchronous state, in a finite network,
is characterized by $r \simeq \frac{1}{\sqrt{N}}$, while 
for $r \equiv 1$ the oscillators are fully synchronized
and intermediate $r$-values correspond to partial synchronization.

Another relevant indicator for the state of the rotator population is the number of
locked oscillators $N_L$, characterized by a vanishingly
small average phase velocity $\bar{\omega_i}\equiv\bar{\frac{d{\theta_i}}{dt}}$, 
and the maximal locking frequency $\Omega_M$, which corresponds to the maximal natural frequency $|\Omega_i|$ 
of the locked oscillators.

In general we will perform sequences of simulations by varying adiabatically
the coupling parameter $K$ with two different protocols.
Namely, for the first protocol (I) the series of simulations is initialized 
for the decoupled system by considering random initial  conditions for $\{\theta_i\}$ 
and $\{\omega_i\}$. Afterwards the coupling is increased in steps
$\Delta K$ until a maximal coupling $K_M$ is reached. For each value
of $K$, apart the very first one, the simulations is initialized by 
employing the last configuration of the previous simulation in the sequence.
For the second protocol (II), starting from the final coupling $K_M$ achieved
by employing the protocol (I), the coupling is reduced in steps $\Delta K$
until $K=0$ is recovered. At each step the system is simulated for a transient time $T_R$ 
followed by a period $T_W$ during which the average value of the order parameter 
${\bar r }$ and  of the velocities $\{ \bar{\omega_i} \}$, as well as $\Omega_M$,
are estimated.

\subsection{Lyapunov Analysis}
\index{maximal Lyapunov exponent}

The stability of Eq.~(\ref{eqPRL}) can be analyzed by following the evolution of infinitesimal perturbations 
$\mathcal{T} = (\delta\dot{\theta}_1,\dots,\delta\dot{\theta}_N,\delta{\theta}_1,\dots,\delta{\theta}_N)$ 
in the tangent space, whose dynamics is ruled by the linearization of 
Eq.~(\ref{eqPRL}) as follows:
\begin{equation}
\label{eq2}
 m \enskip\delta\ddot{\theta}_i + \delta\dot{\theta}_i= 
\frac{K}{N} \sum_{j=1}^N \cos{\left(\theta_j-\theta_i\right)
(\delta\theta_j-\delta\theta_i)} \enskip .
\end{equation}

We will limit to estimate the maximal Lyapunov exponent $\lambda_M$,
by employing the method developed by Benettin {\it et al.}~\cite{benettin1980}.
This amounts to follow the dynamical evolution of the orbit and
of the tangent vector $\mathcal{T}$ for a time lapse $T_W$ by 
performing Gram-Schmidt ortho-normalization at fixed time intervals $\Delta t$, 
after discarding an initial transient evolution $T_R$. 

Furthermore, the values of the components of the maximal Lyapunov
vector ${\cal T}$ can give important information
about the oscillators that are more actively contributing to the chaotic
dynamics. It is useful to introduce the following
squared amplitude component of the normalized vector  for each rotator~\cite{ginelli2011,olmi2015}
\begin{equation}
\xi_i(t) =  [\delta\dot{\theta}_i(t)]^2 
+ [\delta{\theta}_i(t)]^2
\enskip, \qquad i=1,\dots,N \enskip.
\label{loc_vect} 
\end{equation}
 The time average ${\bar \xi_i}$
of this quantity gives a measure of the contribution of each oscillator to the chaotic
dynamics.

\section{Unimodal Frequency Distribution}
\index{unimodal frequency distribution}

\subsection{Hysteretic Synchronization Transitions}
\index{hysteretic transition} \index{synchronization transition}

\begin{figure}[t]
\centering
\includegraphics*[width=.7\textwidth]{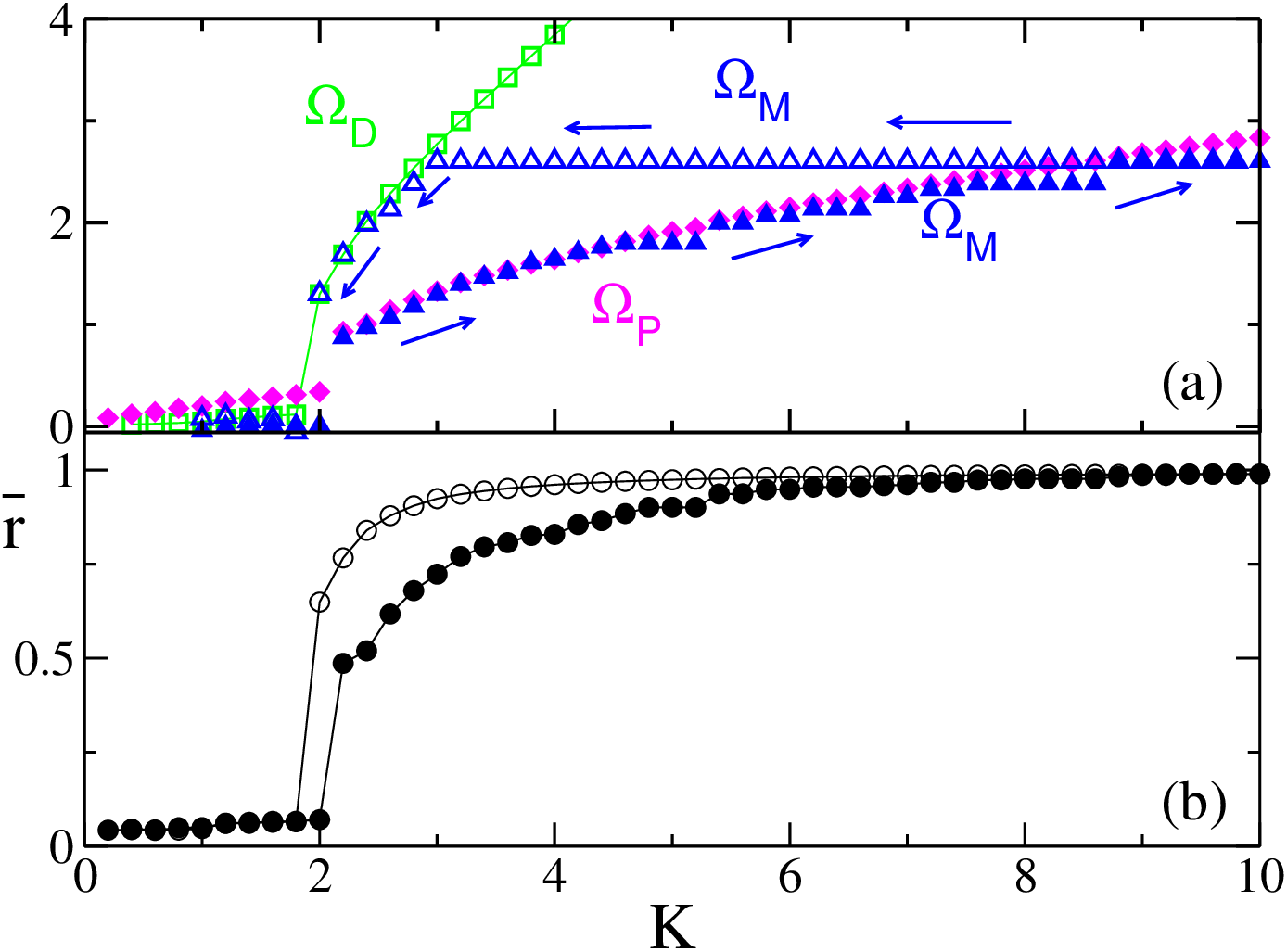}
\caption{Unimodal frequency distribution. (a) Maximal locking frequency $\Omega_M$ (blue triangles)
and (b) time averaged order parameter ${\bar r }$ (black circles) as a function of the coupling $K$
for two series of simulations performed following the protocol (I)
(filled symbols) and the protocol (II) (empty symbols).
The data refer to inertia: $m=2$, for which we set $\Delta K=0.2$
and $K_M =10$; moreover $N=500$, $T_R = 5,000$
and $T_W =200$. The (magenta) diamonds indicate $\Omega_P = \frac{4}{\pi} \sqrt{\frac{K \bar r  }{m}}$
for protocol (I) and the (green) squares $\Omega_D = K \bar r$ for protocol (II).}
\label{omegadis}       
\end{figure}

In Fig.~\ref{omegadis} the results for a sequence of  simulations
obtained by following protocol (I) and (II) are reported 
for a not too small inertia (namely, $m=2$) 
and unimodal frequency distribution. 
For the first protocol the system remains incoherent up to a critical value
$K = K_1^c \simeq 2$, where $\bar r $ jumps to a finite value and then
increases with $K$ reaching $\bar r  \simeq 1$ for sufficiently large 
coupling. Starting from the last state and by reducing $K$ one 
notices that $\bar r $ assumes larger values than during protocol (I)
and the system becomes incoherent at a smaller coupling, namely $K_2^c < K_1^c$.
This is a clear indication of the hysteretic nature of the synchronization transition
in this case. 

For the chosen values of the inertia, we observe the creation of an unique cluster
of $N_L$ locked oscillators with $\bar{\omega_i} \simeq 0$, for larger $m$ the things
will be more complex, as we will discuss in the following. The maximal locking 
frequency $\Omega_M$ becomes finite for $K > K_1^c$ and increases with $K$.
The frequency $\Omega_M$ attains a maximal value when $\bar r  \simeq 1$,
no more oscillators can be recruited in the large locked cluster. 
Once reached this value, even if $K$ is reduced following the protocol (II), 
$\Omega_M$ remains constant for a wide $K$ interval.
Then $\Omega_M$ shows a rapid decrease towards zero by approaching
$K_2^c$. This behavior will be explained in the following two sub-sections.

\subsection{Mean field theory}
\index{mean field theory} \index{inertia}

In order to derive a mean field description of the dynamics of each
single rotator, we can rewrite Eq. (\ref{eqPRL}) by employing the
order parameter definition (\ref{order_parameter}) as follows
\begin{equation}
\label{eqRED}
 m\ddot{\theta}_i + \dot{\theta}_i=\Omega_i -  K r \sin(\theta_i -\phi)
\qquad ;
\end{equation}
this corresponds to the evolution equation for a damped driven pendulum. 
Eq. (\ref{eqRED}) admits, for sufficiently small forcing frequency $\Omega_i$, 
two fixed points: a stable node and a saddle. At larger frequencies $\Omega_i > \Omega_P 
\simeq \frac{4}{\pi}\sqrt{\frac{Kr}{m}}$ the saddle, via a homoclinic bifurcation,
gives rise to a limit cycle. The stable
limit cycle and a stable fixed point coexist until a saddle node 
bifurcation, taking place at $\Omega_i = \Omega_D = K r$, 
leads to the disappearance of the fixed points and for  $\Omega_i  > \Omega_D$ 
only the limit cycle persists. This scenario is correct for sufficiently large inertia;
at small $m$ one has a direct transition from a stable node to
a periodic oscillating orbit at $\Omega_i = \Omega_D = K r$~\cite{strogatz}.
Therefore for sufficiently large $m$ there is a coexistence regime
where, depending on the initial conditions, the single oscillator 
can rotate or stay quiet. The fixed point (limit cycle) solution corresponds to locked
(drifting) rotators.   

The TLO theory~\cite{tanaka1997self,tanaka1997first}
has explained the origin of the first order hysteretic transitions 
by considering two opposite initial states for the network:
(I) the completely incoherent phase ($r=0$) and (II) the completely
synchronized one ($r \equiv 1$). In case (I) the oscillators
are all initially drifting with finite velocities ${\omega_i}$; 
by increasing $K$ the oscillators with smaller 
natural frequencies $|\Omega_i| < \Omega_P$ begin to lock
($\bar{\omega_i}=0$), while the other continue to drift.
This is confirmed by the data reported in Fig.~\ref{omegadis}, 
where it is clear that the locking frequency $\Omega_M$ is
well approximated by $\Omega_P$. The process continues until all the oscillators
are finally locked, leading to $r =1$ and to a plateau in $\Omega_M$.

In the second case, initially all the oscillators are already locked,
with an associated order parameter $r \equiv 1$. 
Therefore, the oscillators can start
to drift only when the stable fixed point solution will disappear, leaving the system 
only with the limit cycle solution. This happens, by decreasing $K$,
whenever $|\Omega_i| \ge \Omega_D = K r$. This is numerically verified,
indeed, as shown in Fig.~\ref{omegadis}, where it is clear that
the maximal locked frequency $\Omega_M$ remains constant
until, by decreasing $K$, it encounters the curve $\Omega_D$ and then
$\Omega_M$ follows this latter curve down towards the asynchronous state.
The case (II) corresponds to the situation observable for the usual Kuramoto model, 
where there is no bistability~\cite{kura}.

In both the considered cases there is a group of desynchronized oscillators
and one of locked oscillators separated by a frequency, $\Omega_P$ ($\Omega_D$ )
in case (I)  (case (II)). At variance with the usual Kuramoto model,
both these groups contribute to the total level of synchronization, namely
 \begin{equation}
\label{rrr}
r = r_L + r_D
\end{equation}
where $r_L$ ($r_D$) is the contribution of the locked (drifting) population.

The contribution of the locked population is simply given by
\begin{equation}
\label{rL}
r_L^{I,II} = K r \int_{-\theta_{P,D}}^{\theta_{P,D}} \cos^2 \theta g(K r \sin \theta) d \theta
\qquad ;
\end{equation}
where $\theta_P = \sin^{-1} (\Omega_P/Kr)$ and $\theta_D =\sin^{-1} (\Omega_D/Kr) \equiv \pi/2$.

The contribution $r_D$ of the drifting rotators is negative and it has been 
estimated analytically by TLO by performing a perturbative expansion
to the fourth order in $1/(m K)$ and $1/(m \Omega)$. 
The obtained expression, valid for sufficiently large inertia, reads as
\begin{equation}
\label{rD}
r_D^{I,II} \simeq - m K r \int_{\Omega_{P,D}}^{\infty}  \frac{1}{(m \Omega)^3} g(\Omega) d \Omega
\qquad ;
\end{equation}
with $g(\Omega) = g(-\Omega)$.

By considering an initially desynchronized (fully synchronized) system 
and by increasing (decreasing) $K$ one can get a theoretical approximation 
for the level of synchronization in the system by employing the 
mean-field expression (\ref{rL}), (\ref{rD}) and (\ref{rrr}) for case I (II). 
In this way, two curves are obtained in the phase plane $(K,r)$, namely
$r^I (K)$ and $r^{II} (K)$. For a certain coupling $K$ the system can attain
all the possible levels of synchronization between $r^I (K)$ and $r^{II} (K)$.

Let us notice that the expression for $r_L$ and $r_D$
reported in Eqs. (\ref{rL}) and (\ref{rD}) are the same 
for case (I) and (II), only the integration
extrema change in the two cases. These are defined by the frequency which
discriminates locked from drifting neuron, that
in case (I) is $\Omega_P$ and in case (II) $\Omega_D$.
It should be noticed that the value of these frequencies is a function
of the order parameter $r$ and of the coupling constant $K$, therefore one
should solve implicit integrals to obtain $r$.
 
\begin{figure}
\centerline{
\includegraphics[angle=0,width=6.cm]{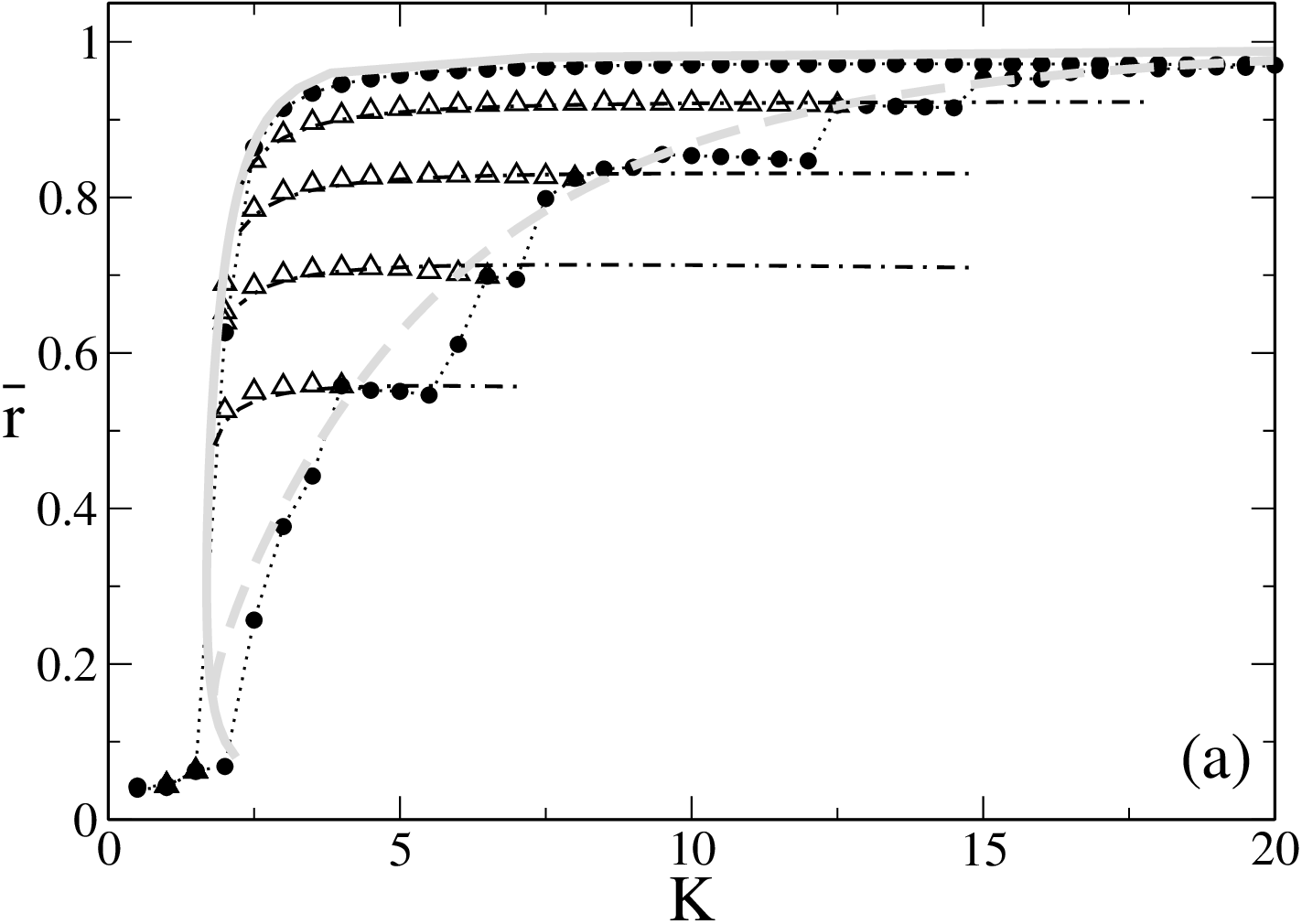}
\includegraphics[angle=0,width=6.1cm]{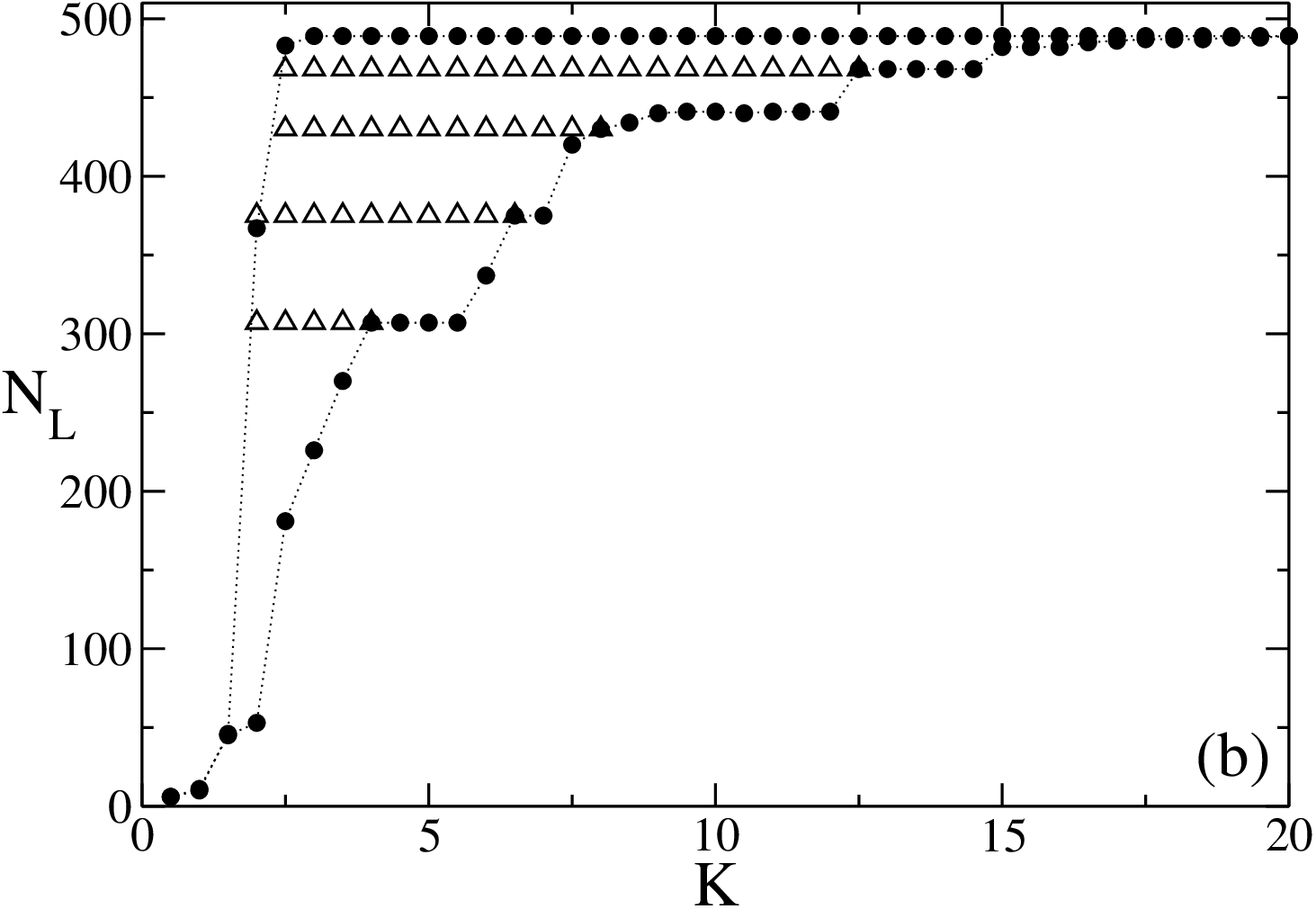}
}
\caption{Unimodal Distribution. 
Panel (a): Average order parameter ${\bar r}$ versus the coupling constant $K$.
Theoretical mean field estimates: the dashed (solid) grey curves refer 
to $r^I  = r^I_L + r^I_D$ ($r^{II}  = r^{II}_L + r^{II}_D$) as obtained by employing
Eqs. (\ref{rL}) and (\ref{rD}) following protocol (I) (protocol (II)); the (black) dot-dashed curves 
are the solutions $r^0(K,\Omega_0)$ of Eq. (\ref{r0}) for different $\Omega_0$ values.
The employed values from bottom to top are:  $\Omega_0=0.79$, 1.09, 1.31 and 1.79. 
Numerical simulations: (black) filled circles have
been obtained by following protocol (I) and then (II) starting from $K=0$ until
$K_M =20$ with steps $\Delta K = 0.5$; (black) empty triangles refer to simulations performed by starting
from a final configuration obtained during protocol (I) and by decreasing the coupling
from such initial configurations. The Panel (b) displays $N_L$ vs $K$ for the numerical simulations
reported in (a). The numerical data refer to $m=6$, $N=500$, $T_R=5000$, and $T_W=200$.
\label{hysteresis}}
\end{figure}

However, one could also fix the discriminating frequency to some
arbitrary value $\Omega_0$ and solve self-consistently the equations
Eqs. (\ref{rrr}), (\ref{rL}), and (\ref{rD}) for 
different values of the coupling $K$. This corresponds to solve the
equation~\index{hysteretic transition}
\begin{equation}
\label{r0}
 \int_{-\theta_{0}}^{\theta_{0}} \cos^2 \theta g(K r^0 \sin \theta) d \theta  
- m \int_{\Omega_{0}}^{\infty}  \frac{1}{(m \Omega)^3} g(\Omega) d \Omega = \frac{1}{K}
\qquad ;
\end{equation}
with $\theta_0 = \sin^{-1} (\Omega_0/Kr^0)$. 
A solution $r^0 = r^0(K,\Omega_0)$ exists provided
that $\Omega_0 \le \Omega_D(K) = r^0 K$.
Therefore the part of the plane delimited by the curve $r^{II}(K)$,
will be filled with the curves $r_0(K)$ obtained for different
$\Omega_0$ values (as shown in Fig.~\ref{hysteresis}(a)).
These solutions represent clusters of $N_L$ oscillators for which the maximal 
locking frequency and $N_L$ do not vary upon changing the coupling strength. 
In particular, for $K > K_2^c$ these states can be observed in
numerical simulations in the portion of the phase space delimited by the two curves 
$r^{I}(K)$ and $r^{II}(K)$ (see Fig.~\ref{hysteresis}(b)).

\subsection{Clusters of Locked and Whirling Oscillators}
\index{locked oscillators} \index{drifting oscillators} \index{standing waves}
 
By observing the results reported in  Fig. \ref{hysteresis}(a) for $m=6$,
it is evident that the numerical data obtained by following 
the procedure (II) are quite well reproduced from the mean field approximation $r^{II}$
(solid grey curve).   
This is not the case for the theoretical estimation $r^I$ (dashed grey curve), which
does not reproduce the step-wise structure revealed for the data corresponding to protocol (I).
This step-wise structure emerges only for sufficiently large inertia 
(as it is clear from Fig.~\ref{omegadis} (b), where it is absent for $m=2$); this is due to the break down 
of the independence of the whirling oscillators: namely, to the formation of 
clusters of drifting oscillators moving coherently 
at the same non zero velocity~\cite{tanaka1997first}.
Oscillators join in small groups to the locked stationary cluster and not 
individually as it happens for smaller inertia; 
this is clearly revealed by the behavior of $N_L$ versus the coupling $K$ 
as reported in Fig.~\ref{hysteresis}(b).

Furthermore, once formed, these stationary locked clusters are particularly
robust, as it can be appreciated by considering 
as initial condition a partially synchronized state obtained following 
protocol (I) for a certain coupling $K_S > K_1$. This state is characterized
by a cluster of $N_L$ locked; if now we reduce the coupling $K$, the
number of locked  oscillators remain constant until we do not reach the 
descending curve obtained with protocol (II), see the black empty triangles in Fig.~\ref{hysteresis}.
On the other hand $\bar r$ decreases slightly with $K$, this behavior
is well reproduced by the mean field solutions of Eq. (\ref{r0}), 
namely $r^0(K,\Omega_0)$ with $\Omega _0 = \Omega_P(K_s,r^I(K_S)) = \frac{4}{\pi} \sqrt{\frac{K_s r^I}{m}}$, 
these are shown in Fig.~\ref{hysteresis} (a) as black dot-dashed lines.
As soon as, by decreasing $K$, the frequency $\Omega_0$ becomes equal or smaller than $\Omega_D$,
the order parameter has a rapid drop towards zero following the upper limit curve $r^{II}$.
These results indicate that hysteretic loops of any size are possible within the 
region delimited by the two curves $\bar r$ obtained by 
following protocol I and II respectively, as shown in \cite{olmi2014}.

\begin{figure}
\centerline{
\includegraphics[angle=0,width=6.1cm]{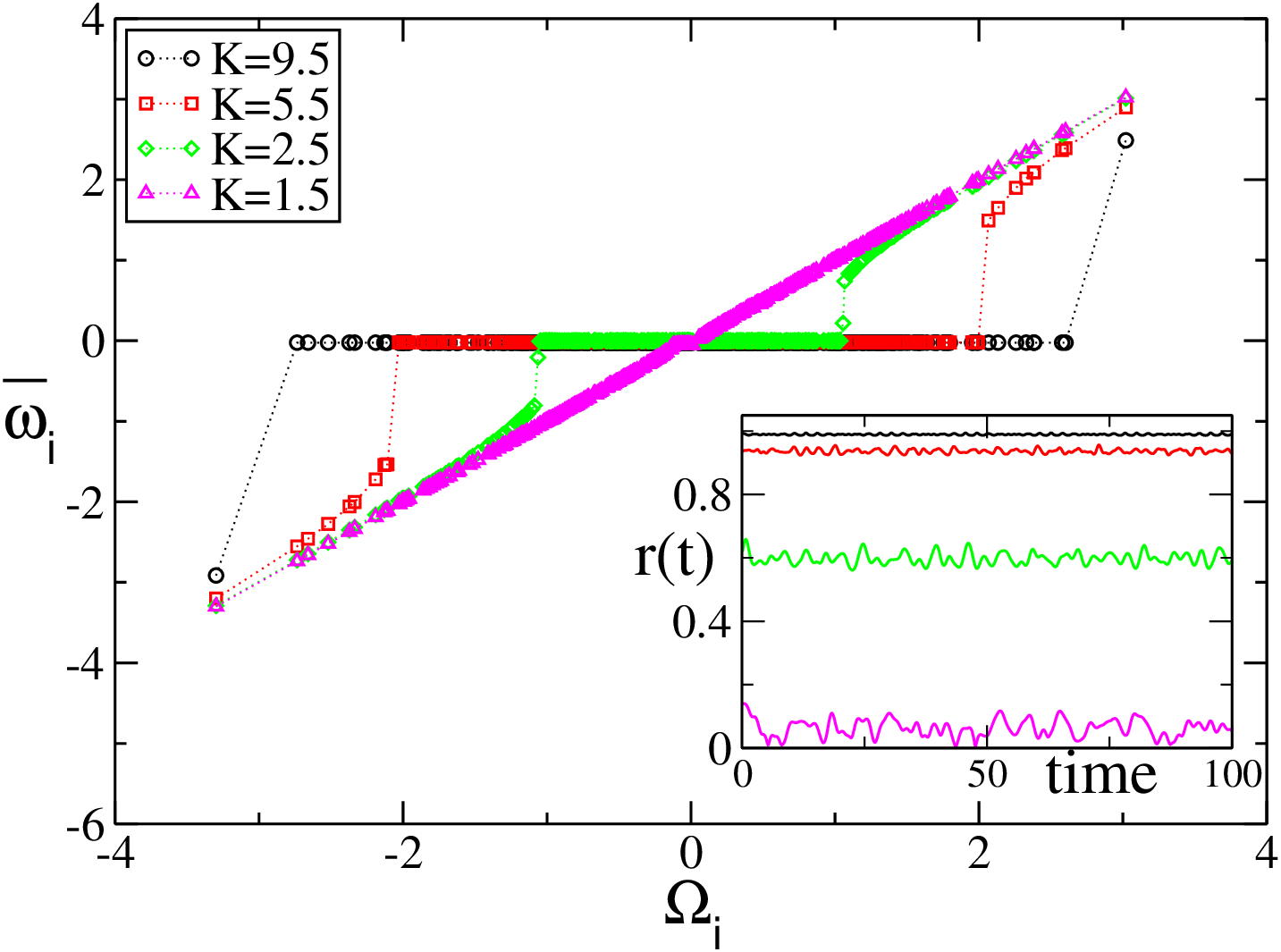}
\includegraphics[angle=0,width=6.cm]{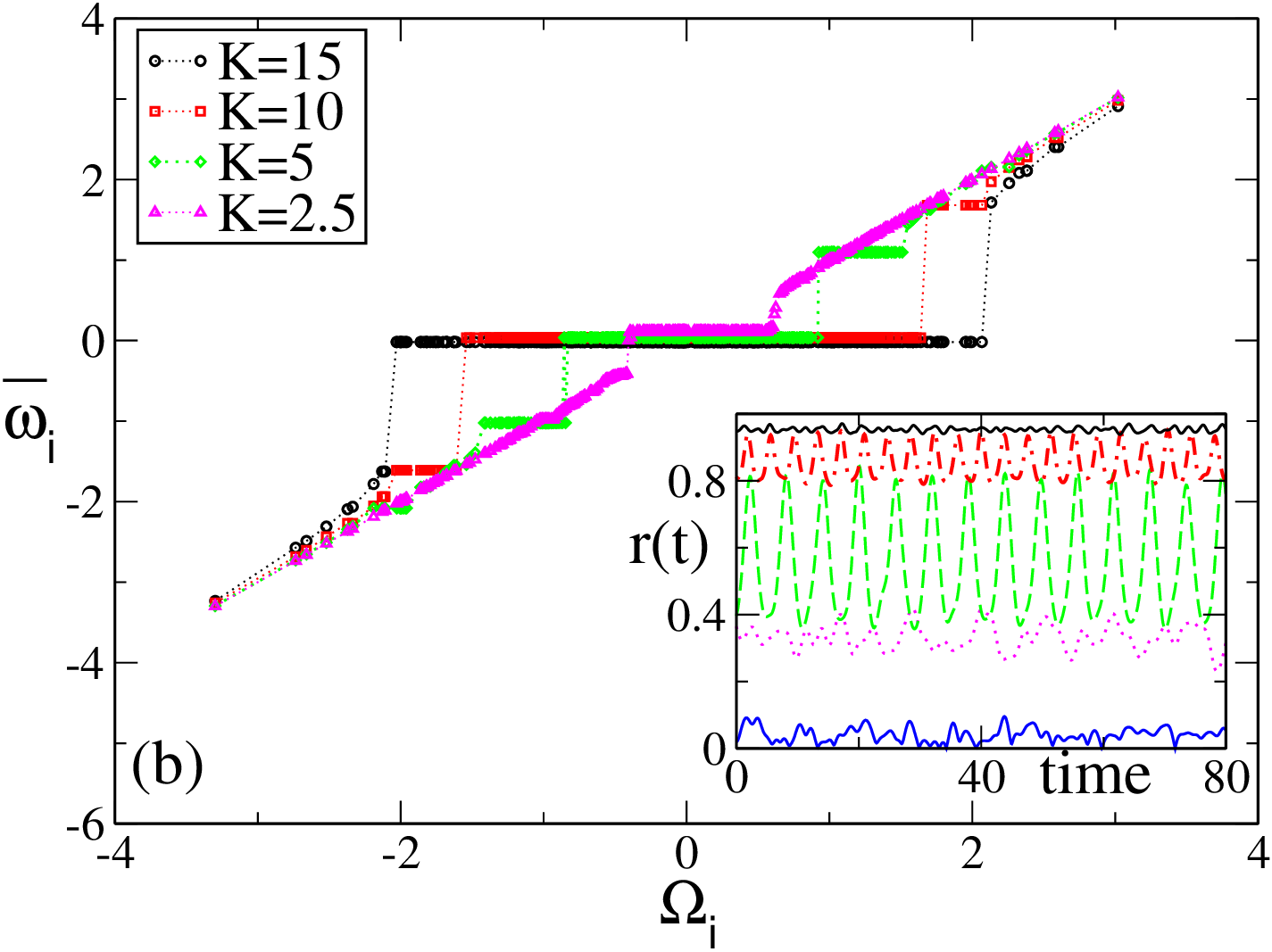}
}
\caption{Unimodal Distribution. Average phase velocity $ \bar{\omega_i}$ of the rotators 
versus their natural frequencies $\Omega_i$ for $N=500$ and inertia $m=2$ (a) and $m=6$ (b).
In panel (a) (panel (b)) magenta triangles refer to $K=1.5$ ($K=2.5$), green diamonds
 to $K=2.5$ ($K=5$), red squares to $K=5.5$ ($K=10$) and black circles to $K=9.5$ ($K=15$). 
The insets report the time evolution of the order parameters $r(t)$
for the corresponding coupling constants, apart for the extra blue line shown 
in the inset in (b) which refers to $K=1$. For each simulation an initial 
transient $T_R \simeq 5,500$ has been discarded and the time
averages have been estimated over a window $T_W=5,000$.  
\label{Plateau}}
\end{figure}

For sufficiently small $m$, the synchronization occurs starting from
the incoherent state via the formation of an unique cluster
of locked oscillators, and the size of this cluster increases with $K$,
as evident from Fig.~\ref{Plateau} (a) for $m=2$. At the same time the
value of $r$ also increases with $K$ and its evolution is characterized only by
finite size fluctuations vanishing in the thermodynamic limit (see the inset of 
Fig.~\ref{Plateau} (a)). As already mentioned, the situation is different 
for sufficiently large inertia, now the partially synchronized phase is characterized 
by the coexistence of the main cluster of locked oscillators with $ \bar{\omega_i} \simeq 0$, 
but also by the emergence of clusters composed by drifting oscillators with common finite 
velocities, see the data for $\bar{\omega_i}$ reported in Fig.~\ref{Plateau} (b) for $m=6$.
In particular, the clusters of whirling oscillators emerge always in couple and they are
characterized by the same average velocity but opposite sign. These states are indicated as
standing waves (SWs), therefore we have a SW coexisting with a partially synchronized
stationary state (PS) (as shown in Fig.~\ref{Plateau} (b).

The effect of these extra clusters on the collective dynamics \index{collective dynamics} is to induce oscillations in the temporal evolution of the order parameter, as one can see from the inset of Fig.~ \ref{Plateau} (b). In presence of drifting clusters characterized by the same average velocity (in absolute value), as for $m=6$ and $K=5$ in Fig.~ \ref{Plateau} (b), 
$r$ exhibits almost regular oscillations and the period of these oscillations corresponds
to the one associated to the oscillators in the drifting cluster.

\section{Bimodal Distribution}
\index{bimodal frequency distribution} \index{inertia}

In this Section we consider a bimodal distribution, we will initially focus on
two almost non overlapping Gaussians, namely we consider $\Omega_0=2$,
while Sub-Section 1.4.2 is devoted to overlapping Gaussians, examined
for $\Omega_0=0.2$.

\subsection{Non Overlapping Gaussians}

\begin{figure}
\centering
\includegraphics[angle=0,width=5.8cm,height=4.4cm]{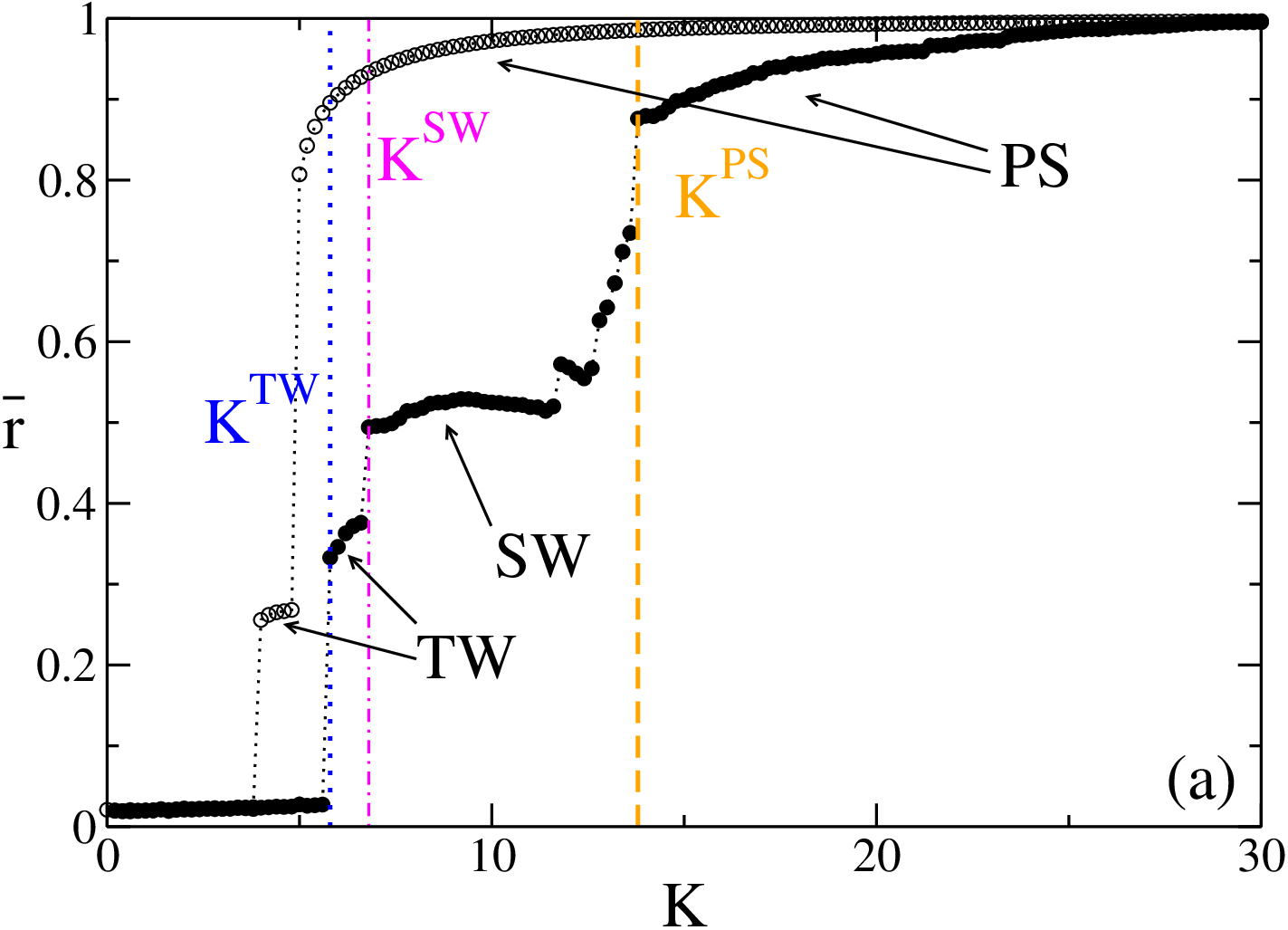} 
\includegraphics[angle=0,width=5.8cm]{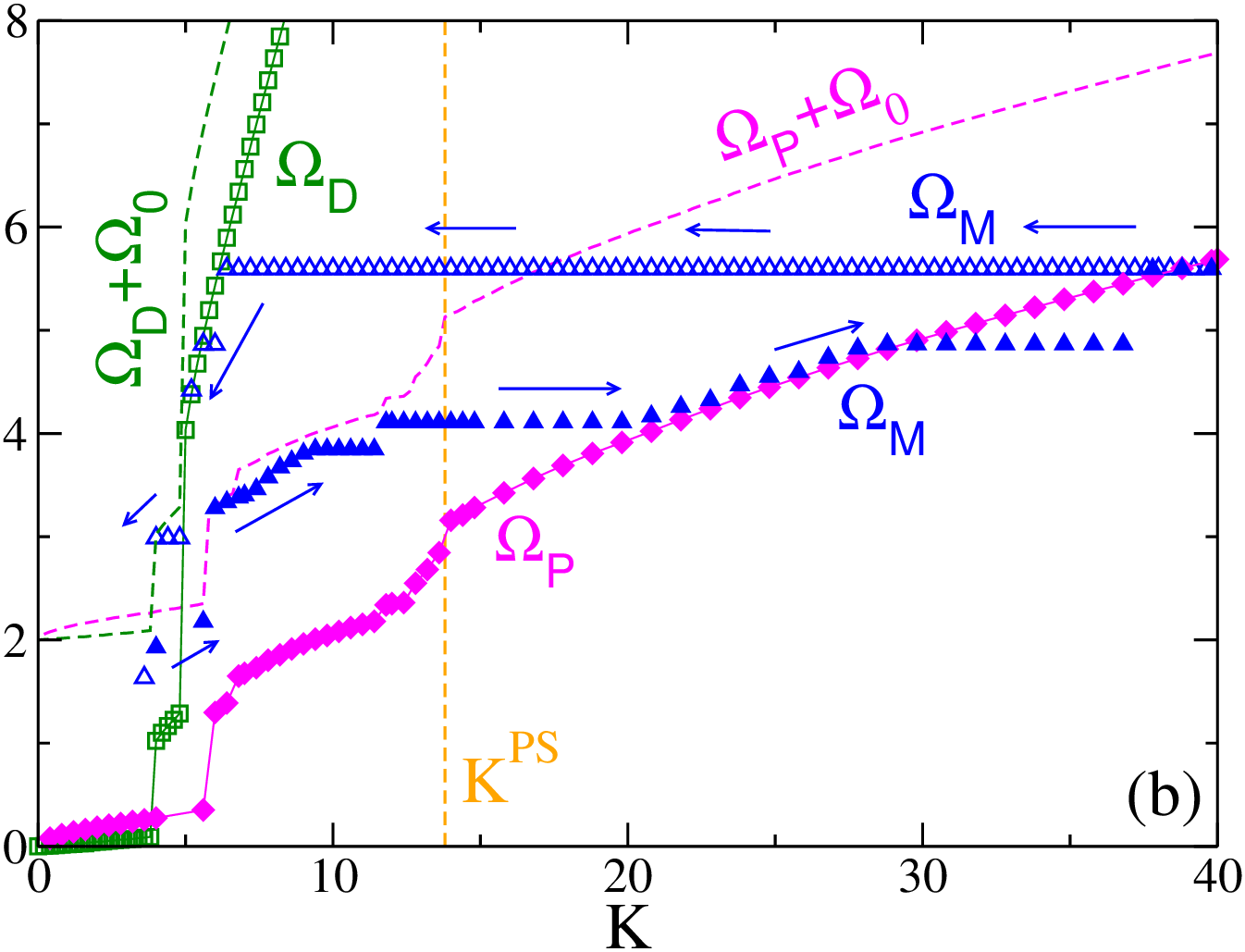} 
\caption{Bimodal frequency distribution. Panel (a):  
Average order parameter ${\bar r}$ versus $K$ for two series of simulations 
performed following the protocol (I) (filled symbols) and (II) (empty symbols).
The dotted vertical blue line refers to $K^{TW}$; the dashed-dotted magenta line to $K^{SW}$.
Panel (b): Maximal locking frequency $\Omega_M$ (blue triangles)
versus $K$ for simulations reported in (a) for protocol (I)
(filled symbols) and (II) (empty symbols).
The magenta diamonds indicate $\Omega_P = \frac{4}{\pi} \sqrt{\frac{K \bar r  }{m}}$
for protocol (I) and the (green) squares $\Omega_D = K \bar r$ for protocol (II).
The dashed magenta line represents the curve $\Omega_P+\Omega_0$ and
 the dashed green curve $\Omega_D+\Omega_0$. 
In both panels the dashed orange vertical line denotes the critical value $K^{PS}$.
The data refer to $m=2$, $\Omega_0=2$, $N=2000$, $T_R = 10,000$ and $T_W =200$,
for the sequence of simulations we employed $\Delta K=0.4$ 
until $K=14.8$ and $\Delta K=0.4$ for $14.8<K<39.8$.
\label{OmegaBim}}
\end{figure}

For $\Omega_0=2$ and sufficiently small inertia ($m = 1$ and 2), we observe
a very rich synchronization transition, as shown in Fig.~\ref{OmegaBim} (a).
In particular, by following protocol (I) we observe that the system leaves the incoherent 
state abruptly by exhibiting a jump to a finite $\bar r$ value at $K^{TW}$;
above such value in the network emerges a single cluster of oscillators, drifting 
together with a finite velocity $\simeq \Omega_0$, this corresponds to Traveling Wave 
(TW) solution~\index{traveling waves}.
By further increasing $K$ a second finite jump of the order
parameter at $K^{SW}$ denotes the passage to a Standing Wave (SW)~\index{standing waves} solution, 
corresponding to two clusters of drifting oscillators with symmetric opposite velocities $\simeq \pm \Omega_0$.
A final jump at $K^{PS}$ leads the system to a Partially Synchronized (PS)~\index{partial synchronization} phase, 
characterized by an unique cluster of locked rotators~\index{locked oscillators} with zero average velocity. 
By increasing the coupling the PS state smoothly approaches the fully synchronized regime~\index{full synchrony}. 
Starting from this final state the return sequence of simulations,
following protocol (II), displays a simpler phenomenology. 
The network stays in the PS regime, characterized by an order parameter
larger than that measured during protocol (I) simulations, until
$K^{DS} < K^{TW}$.  For smaller coupling, the system leaves the PS state;
however, depending on the realization of the natural frequencies and on 
the initial conditions, it can ends up or in a TW (most of the cases) or in 
a SW, or it can even reach directly the incoherent state
(as shown in Fig.~\ref{OmegaBim} (a) and Fig.~\ref{rvsK_bim} (a)).
This first analysis clearly shows  hysteretic effects and coexistence
of macroscopic states with different level of synchronization 
for a wide range of couplings.

Let us now try to examine the observed transitions in terms of
the maximal locking frequency, $\Omega_M$. 
This frequency is now defined in a different way with respect to the
unimodal distribution, in the present case $\Omega_M$ represents the maximal absolute
value of the natural frequencies of the oscillators belonging to the main
clusters present in the system, therefore in this estimation
are considered both stationary and drifting clusters.
As shown in Fig.~\ref{OmegaBim} (b), $\Omega_M$ increases with $K$ for protocol (I)
simulations. In particular $\Omega_M$ shows a finite jump in correspondence of $K=K^{TW}$,
and then its evolution is reasonably well approximated by the curve
$\Omega_P+\Omega_0$, where $\Omega_P = \frac{4}{\pi} \sqrt{\frac{K \bar r  }{m}}$.
By approaching $K^{PS}$ the maximal frequency displays a constant plateau
which extends beyond $K^{PS}$, this indicates that the two symmetric
drifting clusters merge at $K=K^{PS}$ giving rise to an unique locked cluster
with zero average velocity, however no other oscillators join this cluster
up to a larger coupling. Whenever this happens , $\Omega_M$ starts again to increase,
but this time it follows the curve $\Omega_P = \frac{4}{\pi} \sqrt{\frac{K \bar r  }{m}}$.
Finally, for $\bar{r}\simeq 1$ the maximal locking frequency attains a maximal value. Moreover, 
by reducing the coupling, following now protocol (II), $\Omega_M$ remains stacked to 
such a value for a wide $K$ interval. The fully synchronized cluster is difficult to
break down due to the inertia effects. Finally, $\Omega_M$ reveals a rapid decrease 
towards zero whenever it encounters the curve $\Omega_D= K \bar r$, initially it follows
this curve, however as soon as the system desynchronizes towards a TW
the decrease of $\Omega_M$ is better described by the curve $\Omega_D+\Omega_0$.
The observed behavior can be explained by the fact that for $K<K^{PS}$
($K < K^{(DS)}$) for protocol (I) (protocol (II)),
the network behaves as two independent sub-networks each characterized
by an unimodal frequency distribution, one centered at $\Omega_0$ and the other
one at $-\Omega_0$. The extension of the analysis reported in Section 1.3.2
for an unimodal distribution not centered around zero simply amounts to shift the 
limiting curves  $\Omega_P$ and $\Omega_D$ by $\Omega_0$. However,
for sufficiently large coupling constant, once the system exhibits only
one large cluster with zero velocity, the network behaves as 
a single entity and $\Omega_M$ closely follows $\Omega_P$ or $\Omega_D$
as for a single unimodal distribution centered in zero.

\begin{figure}[t]
\centering
\includegraphics*[width=.9\textwidth]{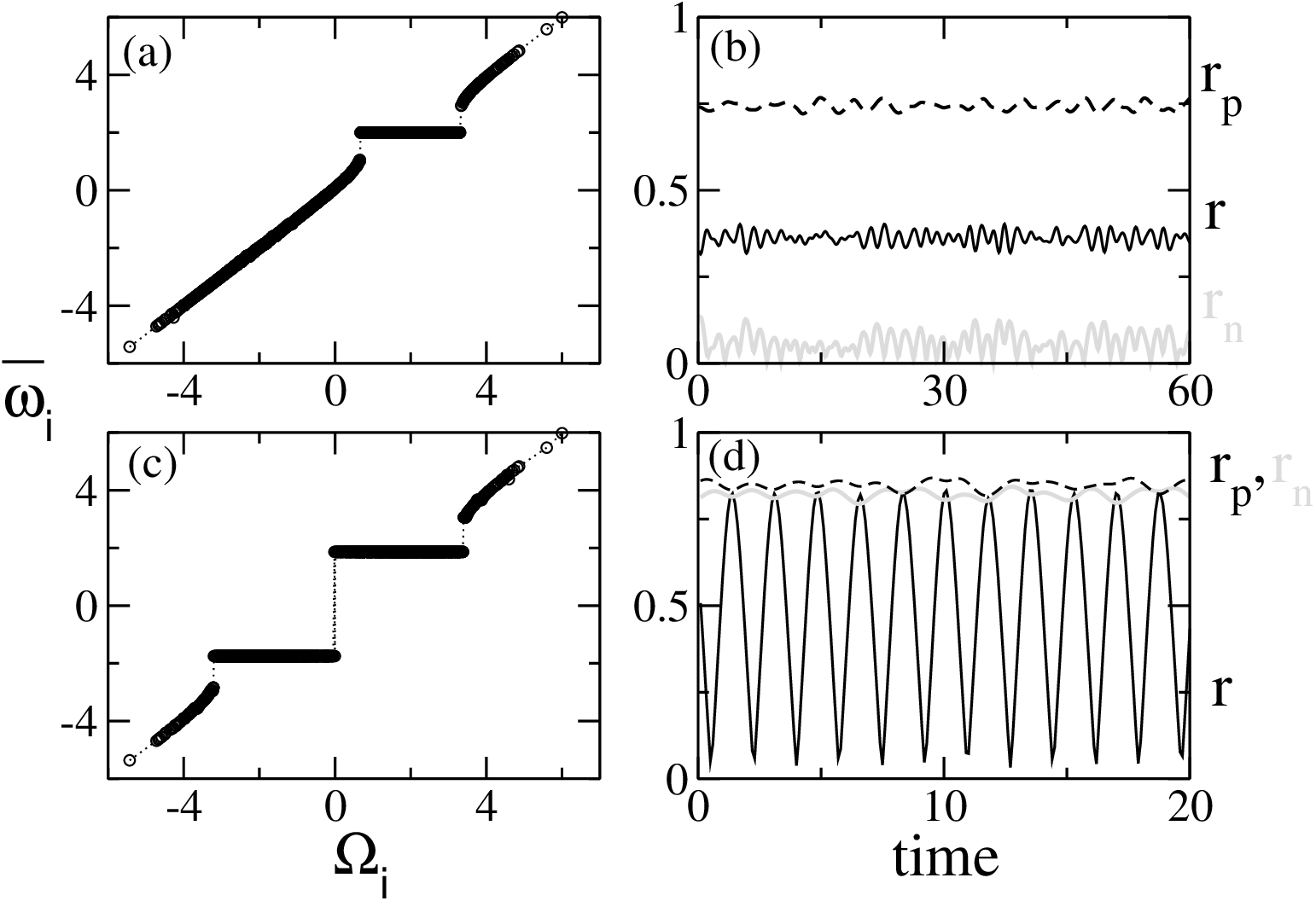}
\caption{Bimodal frequency distributions. Average phase velocity $ \bar{\omega_i}$ of the 
rotators versus their natural frequencies $\Omega_i$ for coupling strength K=6.2 (a)
and K=6.8 (c). Panels (b) and (d) display the order parameter $r(t)$ (black line) 
versus time for the same coupling constants as in (a) and (c), respectively.
The dashed black (continuous grey) line denotes the time evolution of $r_p$ ($r_n$).
For each simulation  an initial transient time $T_R =1000$ has been discarded
and the average are estimated over a time interval $T_W=200$.
In both cases $m=2$, $N=2000$ and $\Omega_0=2$. 
}
\label{fig6}       
\end{figure}

\begin{figure}
\includegraphics[angle=0,width=6.1cm]{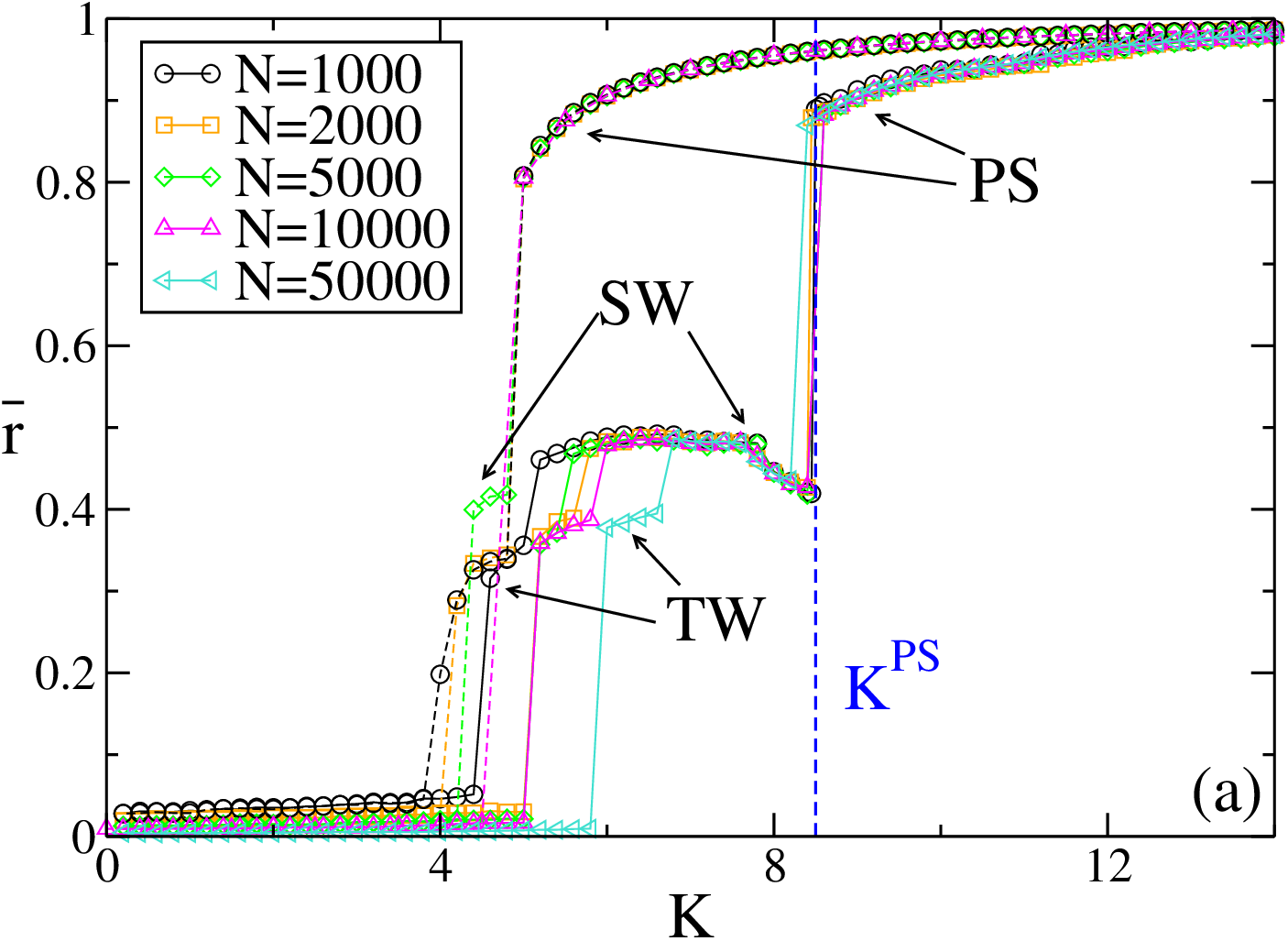}
\includegraphics[angle=0,width=6.cm]{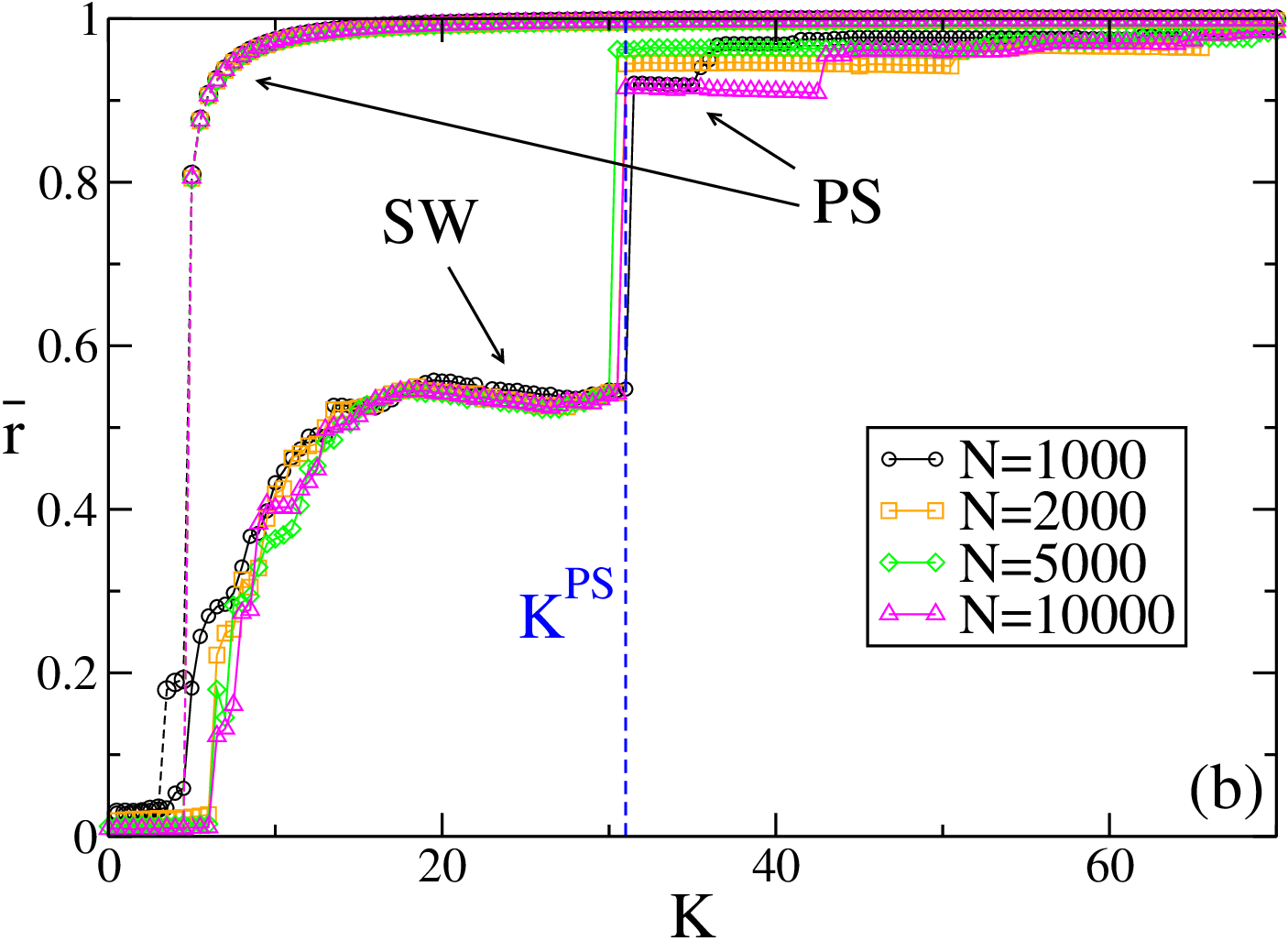} 
\caption{Bimodal frequency distribution with $\Omega_0=2$. Average order parameter ${\bar r}$ versus 
$K$ for various system sizes $N$: (a) $m=1$, (b) $m=6$. The numerical data have been obtained by following 
protocol (I) and then protocol (II) from $K=0$ up to $K_M=20$ ($K_M=200$) for inertia m=1 (m=6) with
$\Delta K=0.2$ ($\Delta K=0.5$). The vertical dashed blue line refers to $K^{PS}$. 
The average has been performed over a time window $T_W=200$, after discarding a transient time
$T_R=5,000 - 50,000$ depending on the system size; 
the larger $T_R$ have been employed for the larger N.
\label{rvsK_bim}}
\end{figure}

Let us now describe the TW and SW states in more details with the help of the 
examples reported in Fig.~\ref{fig6} for inertia $m=2$ and $N=2000$.
The TW is an  asymmetrical cluster of whirling oscillators with a finite velocity
$ \bar{\omega_i}\simeq \Omega_0$, in particular in Fig.~\ref{fig6} (a)
the oscillators have natural frequencies in a range around $\Omega_0$, 
namely $0.67 \le \Omega_i \le 3.34$. The effect of this cluster 
on the collective dynamics is to increase the average value of the order parameter 
without inducing any clear oscillating behavior in $r(t)$.
However, oscillators with positive natural frequencies are much more synchronized with 
respect to the ones with negative frequencies, as can 
be inferred by observing the order parameter $r_p$ ($r_n$) estimated only
on the the sub-population of oscillators with positive (negative) natural frequencies
and reported in Fig.~\ref{fig6} (b).
We believe that the emergence of the TW state is related to the finite sampling of the distribution 
of the natural frequencies, which due to finite size effects can be non perfectly symmetric.
The asymmetric cluster will emerge around $+\Omega_0$ ($-\Omega_0$)
depending on the positive (negative) sign of the average natural frequency.

The SWs are observable at larger coupling constants,
this is characterized by two symmetrical clusters with
opposite average velocities $\simeq \pm \Omega_0$, as shown in Fig. \ref{fig6} (c).
The presence of these two clusters now induces clear periodic oscillations
in the order parameter $r(t)$, as observable in Fig. \ref{fig6} (d).
The period of the oscillations is related to $|\Omega_0|$, i.e. the average frequency
of the clustered oscillators. However, at variance with the results reported in
Fig.~\ref{Plateau} (b) for the unimodal distribution the two symmetric clusters do not coexist
with a cluster of locked oscillators with zero average velocity.
By examining separately $r_p$ and $r_n$ reported in Fig.~\ref{Plateau} (d), we notice that 
each sub-population is much more synchronized than the global one, 
in fact $r_p$ and $r_n$ have a higher average value than $r$ with superimposed
irregular oscillations. \index{collective dynamics}

The data reported so far refer to a single system size, however finite size effects \index{finite size effects}
are quite relevant for this model, as shown in \cite{olmi2014} for unimodal distributions.
In Fig.~ \ref{rvsK_bim} (a) we report the synchronization transition for 
several system sizes, namely $1,000 \le N \le 50,000$ for a small inertia value ($m=1$).
We observe that $K^{TW}$ and $K^{SW}$ increase with the size $N$; in particular,
the incoherent state is observable on a wider coupling interval by increasing $N$
(similarly to what reported in \cite{olmi2014} for unimodal distributions).
Finite size fluctuations induce transitions from the incoherent branch
to the TW branch and from this to the SW branch. The fact that we
do not observe transition back to the original states indicates that the
energy barriers are higher from these sides. A quite astonishing result 
is the fact that the transition value $K^{PS}$ and $K^{DS}$ seem 
completely independent from $N$. The combination of these 
results seem to suggest that in the thermodynamic limit the
incoherent state will loose stability at $K^{PS}$ and therefore
the two branches corresponding to TW and SW will be no more visited,
at least by following protocol (I).

By observing all the data reported in Fig.~ \ref{rvsK_bim} (a) for various $N$
and for protocol (I) and (II), it seems that there are clear indications that
the two branches of solutions, corresponding to TW and SW, emerge 
via a supercritical bifurcation at the same coupling, namely $K \simeq 3.8$, 
while the transition to PS is clearly subcritical.
These results confirm the analysis reported in~\cite{acebron2000} for
a system with noise (in particular, see Fig. 17 in that paper). 
However, Acebr\'on {\it et al.} affirm that the SW is stable, while
the TW is unstable. From our results, both branches seem to become inaccessible 
(in absence of noise) from the incoherent state, while at least a part of these branches appear to
be reachable from the PS state by decreasing $K$ below $K^{DS}$ following
protocol (II). Another important 
difference with respect to the results reported in~\cite{acebron2000}
is that the PS regime is clearly hysteretic revealing two coexisting branches of PS
states visited by following protocol (I) or (II).

As shown in  Fig. \ref{rvsK_bim} (b),
for larger inertia (namely, $m=6$), the transition from the incoherent state 
following protocol (I) occurs via the emergence of many small clusters
leading finally to a SW state. In this case the critical value at which 
the incoherent state looses stability seems to saturate to a constant value
already for $N \ge 2,000$. The value of $K^{PS}$ is also in this case insensible
to the system size. For large inertia values, the TWs seem no more observable.

\begin{figure}[t]
\centering
\includegraphics*[width=.5\textwidth]{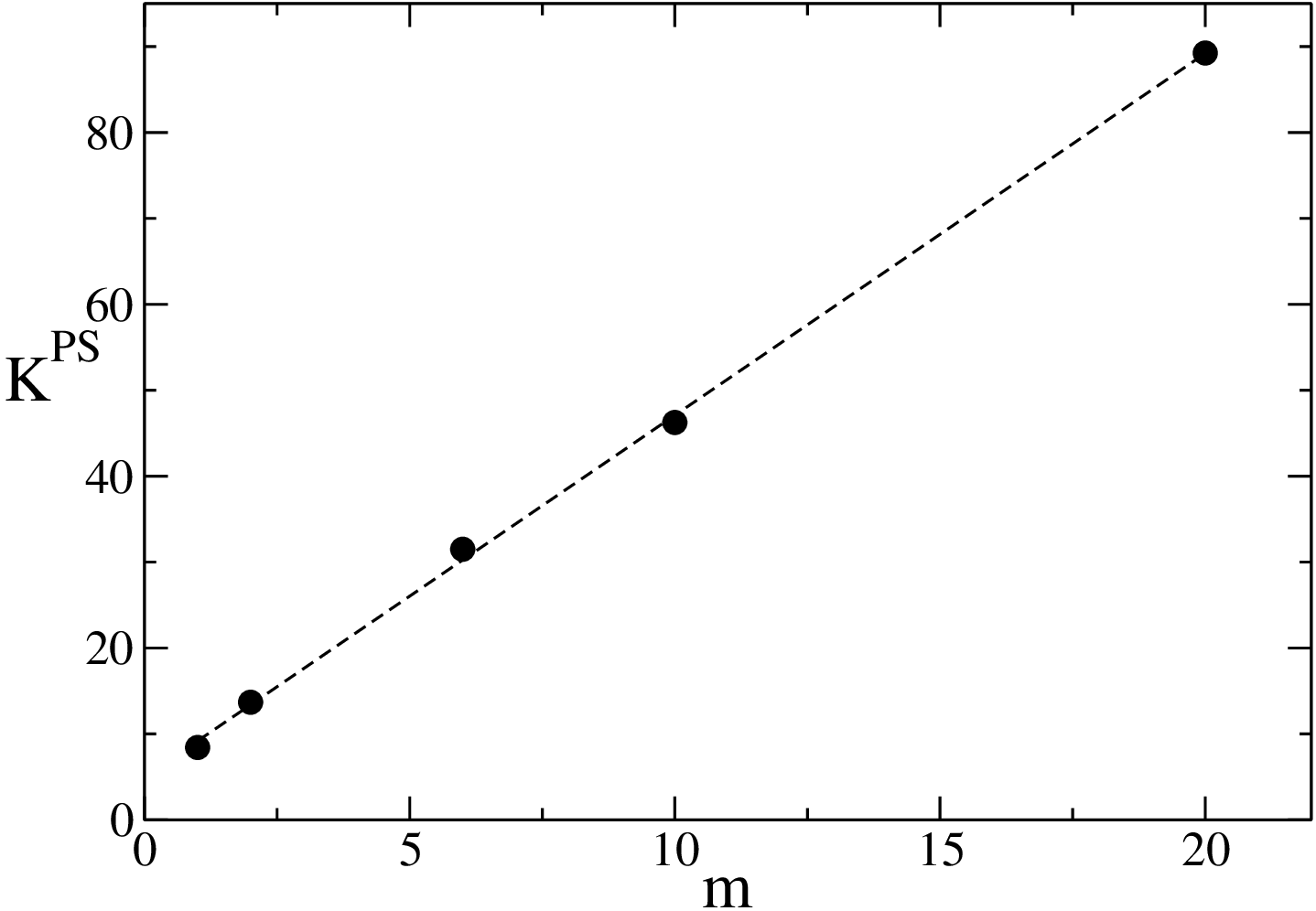}
\caption{Critical value $K^{PS}$ as a function of the inertia.
The dashed line represents the fit of the numerical data and indicates a linear
increasing of $K^{PS}$ as a function of the inertia being the fit $K^{PS}=4(1.245 + 1.0525 m)$. For all cases 
$N=2000$, $\Omega_0=2$. The data have been obtained by employing protocol (I) and for each simulation  
an initial transient time $T_R =5000$ has been discarded and data are averaged over a time $T_W=200$.  }
\label{KvsMass}       
\end{figure}

As a further aspect, we will report the numerical results of the dependence on the 
inertia of the critical coupling constant $K^{PS}$, while the value of $K^{DS} \simeq 4.9$ 
is independent not only by $N$, but also by the inertia. 
As shown in Fig. \ref{KvsMass}, $K^{PS}$ increases linearly with the inertia and
this scaling is already valid for not too large inertia values. 
The linear scaling with the inertia is analogous to the scaling recently found
within a theoretical mean-field analysis for the coupling $K_1^{MF}$, which delimits the range of 
linear stability of the asynchronous state~\cite{acebron2000,gupta2014}.
In particular, the authors in~\cite{olmi2014} have shown for a Gaussian unimodal distribution of width
$\sigma$ that $K_1^{MF} \simeq 2 \sigma (0.64 + m \sigma)$, which shows a linear dependence on the inertia 
and a quadratic dependence on the variance of the frequency distribution.

\begin{figure}
\includegraphics[angle=0,width=6.1cm]{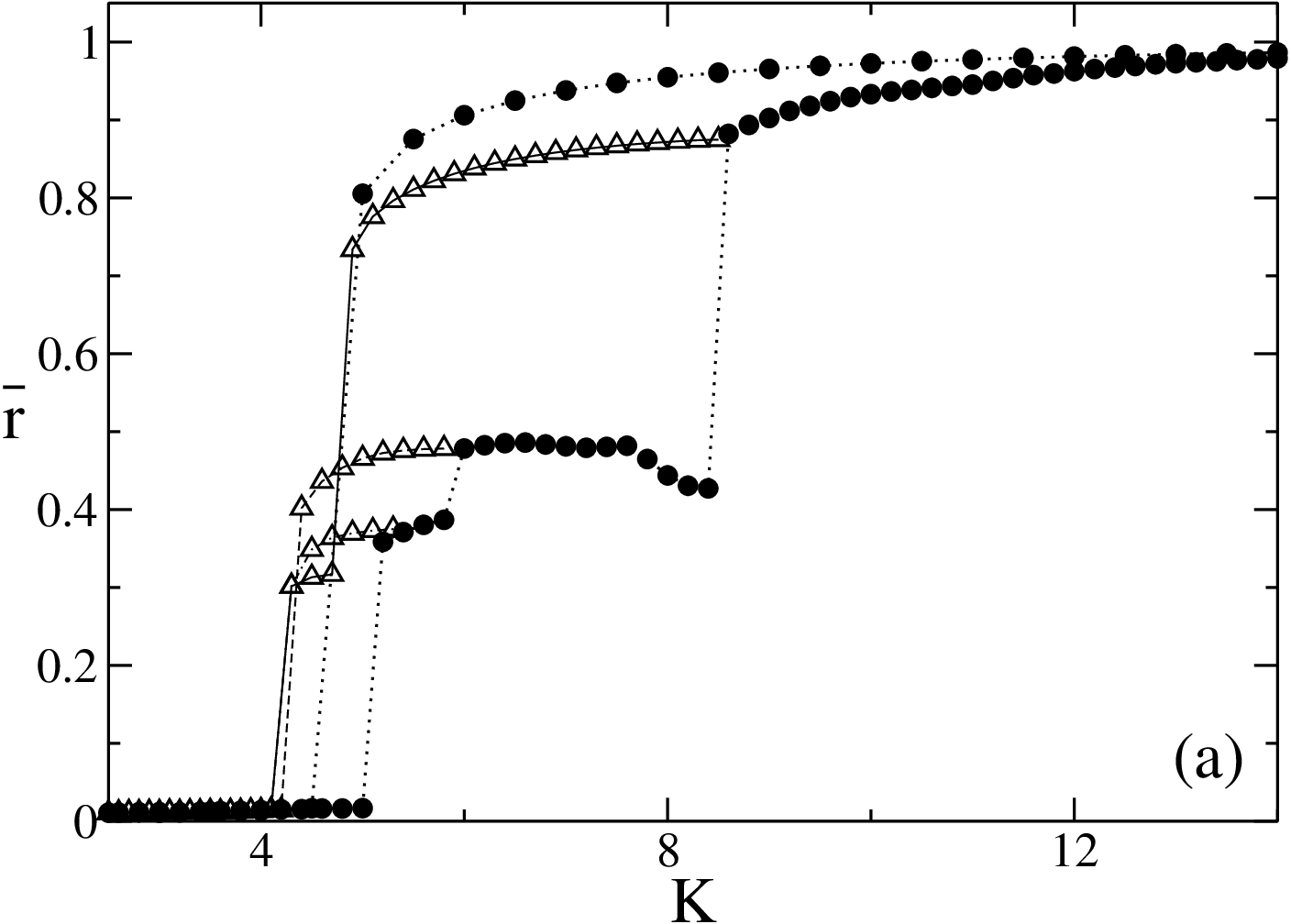}
\includegraphics[angle=0,width=6.cm]{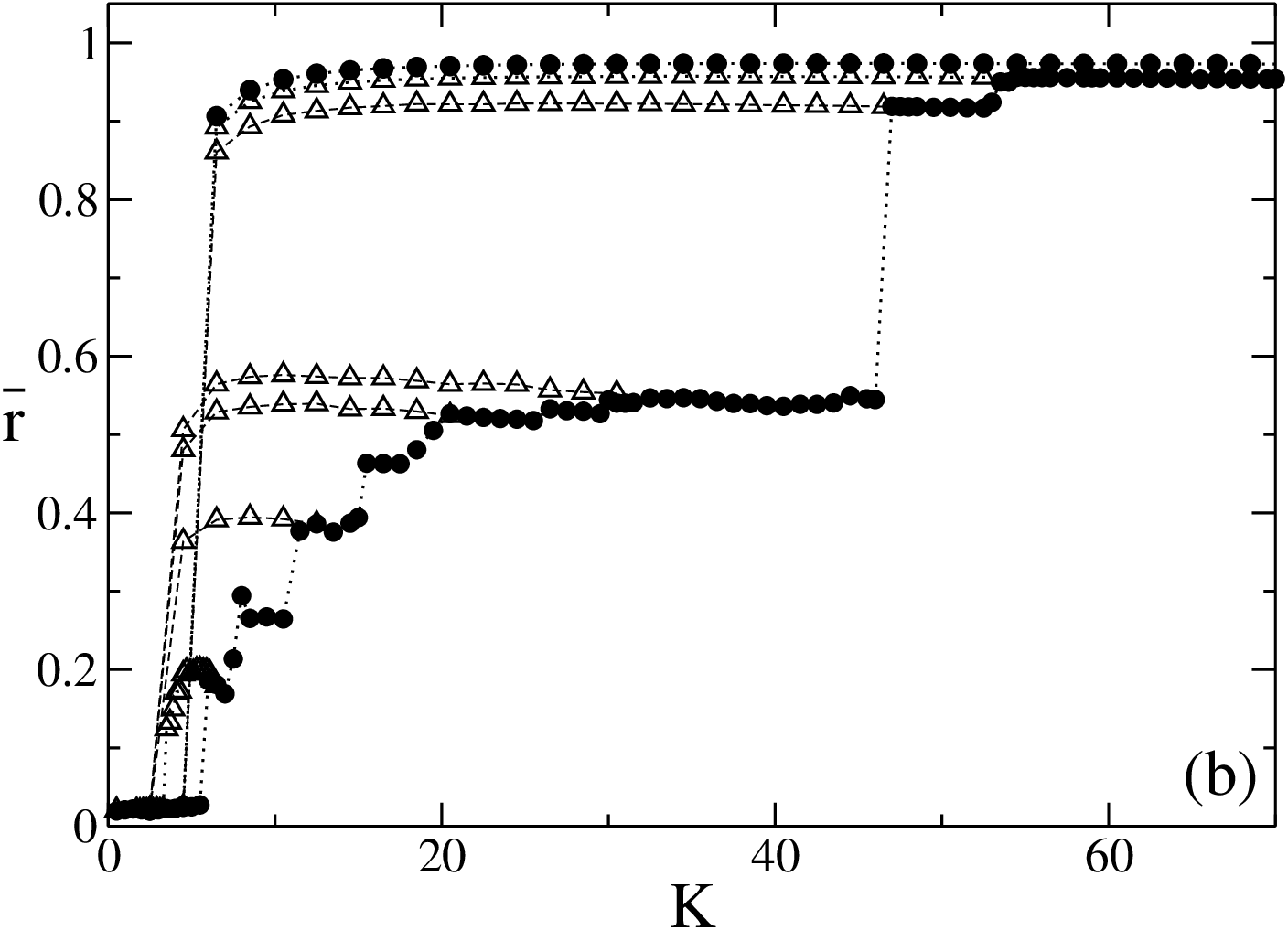} 
\caption{Bimodal frequency distribution.
Average order parameter ${\bar r}$ versus the coupling constant $K$
for $m=1$ and $N=10,000$ (panel (a)) and for  $m=10$ and $N=2,000$ (panel (b)). The
filled circles have been obtained by following protocol (I) and then (II) starting from $K=0$ until 
$K_M =20$ ($K_M =100$) with steps $\Delta K = 0.2$ ($\Delta K = 0.5$); 
the empty triangles refer to simulations performed by starting
from a final configuration obtained during protocol (I) and by decreasing the coupling
from such initial configurations. 
The numerical data refer to $\Omega_0=2$, $T_R=50000$ ($T_R=5000$), and $T_W=2000$.
\label{isteresi_bim}}
\end{figure}

In the final part of this sub-section we perform an analysis
analogous to that reported in Sub-Sect. 1.3.3, in particular
starting from states with a finite level of synchronization
obtained by following protocol (I) we decrease the coupling
and observe how these states evolve. 
In Fig.~ \ref{isteresi_bim}, we report the results of these
simulations (shown as empty triangles) for two different inertia values,
namely $m=1$ and $m=10$. Starting from PS states we observe that
the cluster survives until the descending curve obtained with protocol (II) is encountered, 
analogously to the results reported in Fig.~\ref{Plateau} for the unimodal
distribution. Therefore any part of the hysteretic portion of the $(K,r)$-plane 
delimited by the PS curves obtained via protocol (I) or (II) is accessible~\index{hysteresis}.
However, if one starts for $m=1$ from a TW or a SW state, one observes
only two curves (corresponding to the TW and SW branches previously
discussed) which seem to end up at the same critical coupling which is smaller than $K^{DS}$. 
Therefore it seems that there are no evidences of hysteresis 
for this small inertia for SW and TW solutions (as shown in Fig.~ \ref{isteresi_bim} (a)).
For large inertia values $m=10$, since now, apart the SW solutions, there are solutions
with many small clusters, the situations is much more complex. By starting from different
values of $K < K^{PS}$ and by decreasing $K$, these curves seem all to end up at the
same critical coupling smaller $K^{PS}$, see Fig.~ \ref{isteresi_bim} (b). These
results suggest that for large inertia values is possible to observe a continuum 
of possible states even starting from states characterized by (many) drifting clusters
and that these states coexist in a wide range of coupling.

\subsection{Overlapping Gaussians}

\begin{figure}
\includegraphics[angle=0,width=6.1cm]{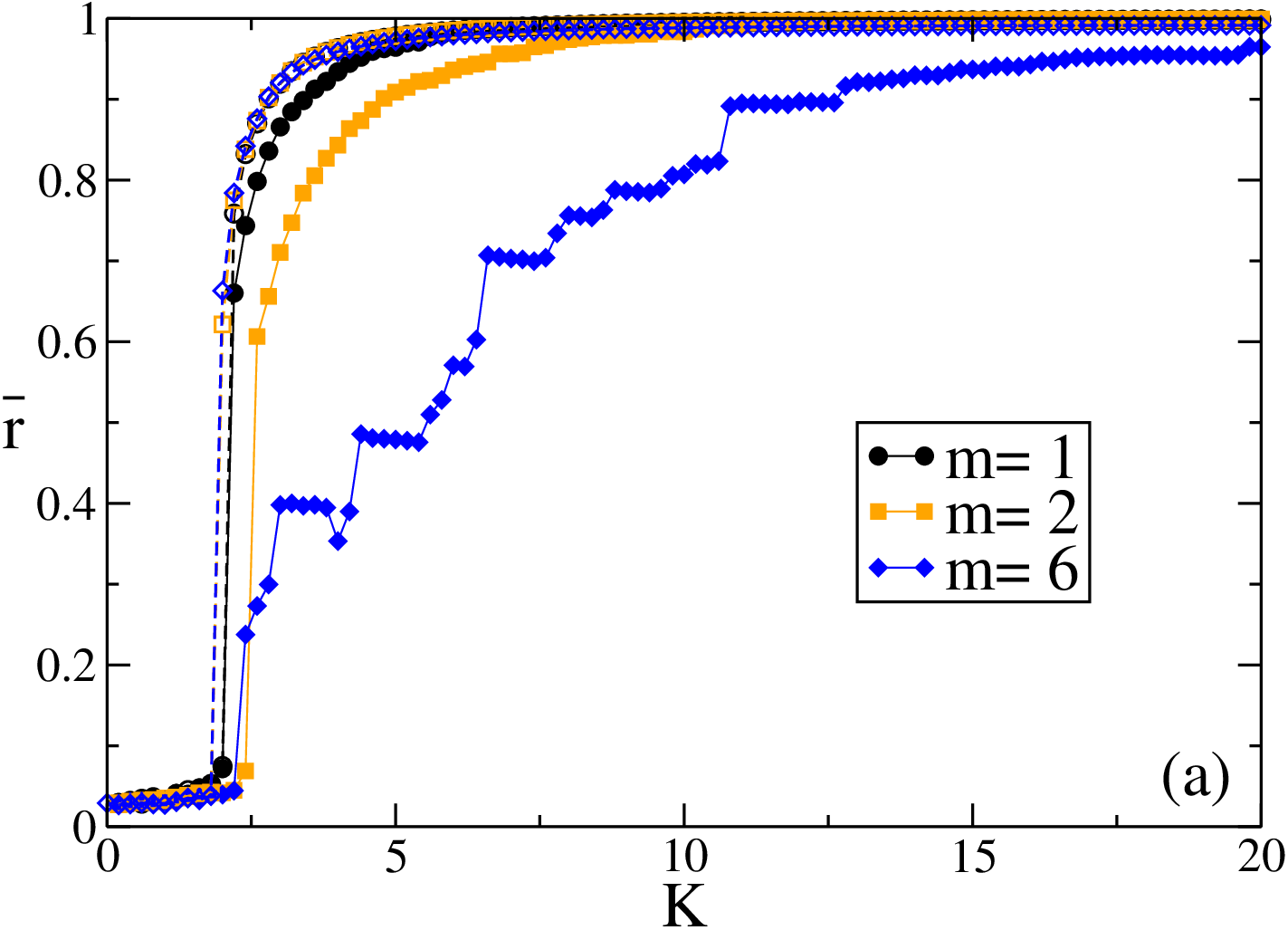}
\includegraphics[angle=0,width=6.cm]{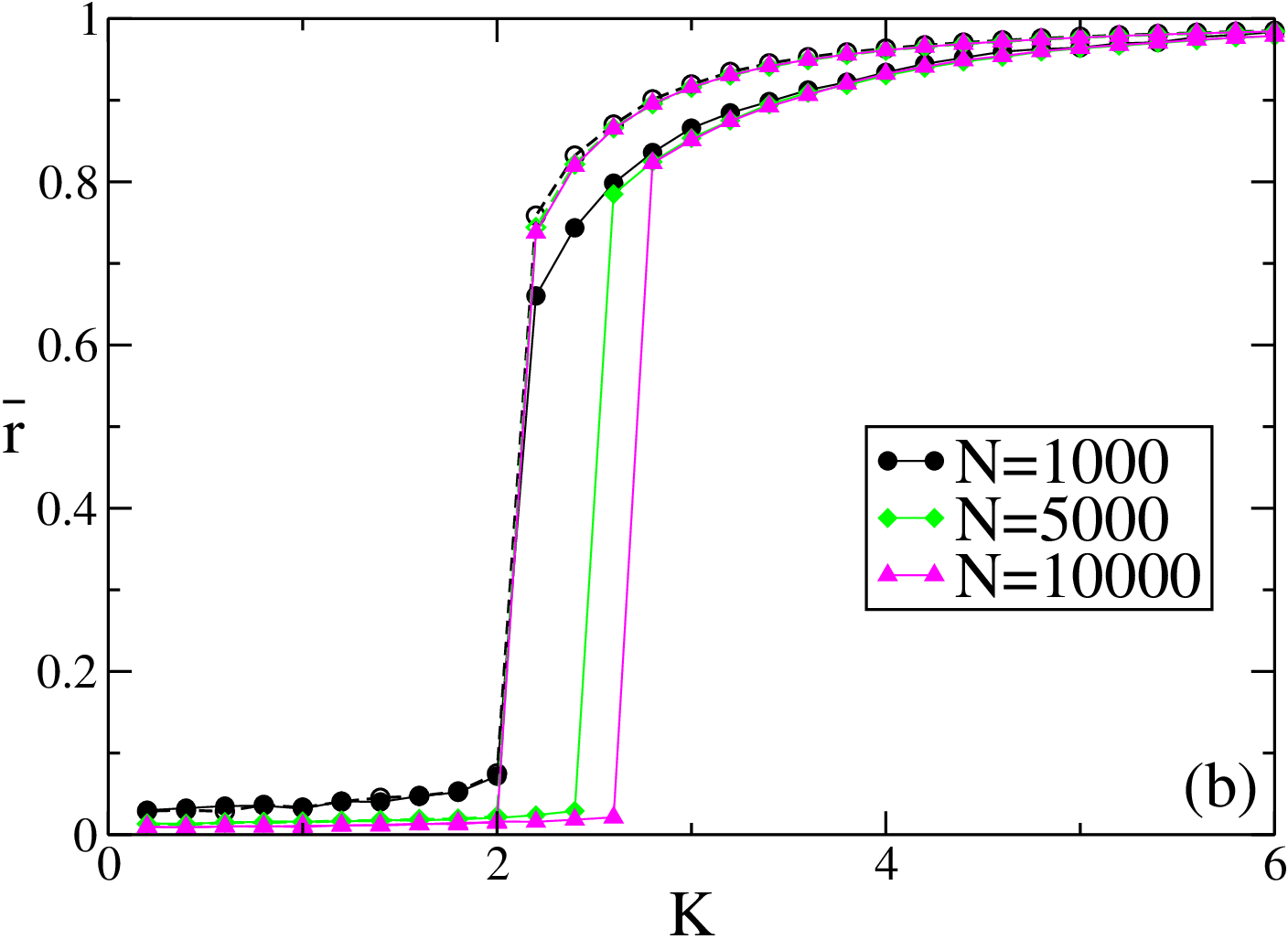} 
\caption{Bimodal frequency distribution for $\Omega_0=0.2$.
Panel (a) : Average order parameter ${\bar r}$ versus $K$
for various inertia values and N=1000. 
The numerical data have been obtained by following protocol (I) and then protocol 
(II) from $K=0$ up to $K_M=20$ for all inertia values with $\Delta K=0.2$ .
Panel (b): Average order parameter ${\bar r}$ versus the coupling constant $K$
for various system sizes N and m=1. Data have been obtained by averaging the order parameter over a 
time window $T_W=200$, after discarding a transient time $T_R=5,000 - 50,000$ depending on the system size.   
\label{rvsK_omega02}}
\end{figure}

In this sub-section, we analyze a bimodal distribution, where the two
Gaussians are largely overlapping, since $\Omega_0=0.2$.
In this case we expect to observe a phenomenology of the synchronization
transition quite similar to the one seen for the unimodal case.
In Fig. \ref{rvsK_omega02}(a) is reported the average order parameter ${\bar r}$ 
versus the coupling constant $K$ estimated by following
protocol (I) and (II) for various inertia values and for a fixed system size, namely $N=10,000$.
We observe that all the curves obtained for protocol (II) almost overlap
irrespectively of the used inertia, while the protocol (I) curves reveal a strong dependence 
on $m$. In particular, the hysteretic region widens with $m$.
For small inertia values, namely $m=1$ and 2, there is a sudden transition from the
asynchronous state to a PS state at $K^{PS}$ and neither traveling waves nor standing waves are observable: 
a single cluster at zero velocity emerges in correspondence of $K^{PS}$ and the order parameter 
never shows oscillating behavior in time.

For $m=6$, it is possible to observe a scenario similar to the one reported in Fig. \ref{Plateau} (b),
where not only a cluster at zero velocity is present, but also two symmetrical clusters at finite velocities emerge. 
In particular, following protocol (I) for $K > 2.4$ a small cluster of locked oscillators emerges; at larger coupling,
namely $K\geq 3$, two symmetrical clusters of whirling oscillators emerge and coexist
with the zero velocity cluster. Finally, at $K=10.8$ the PS regime arises,
corresponding to a single large cluster of locked oscillators. Furthermore,
in the range $3 \leq K < 10.8$ the order parameter reveals irregular oscillations.
A more detailed analysis is needed to understand the origin
of these oscillations as done in the next Section.

Finally, we examine the influence of the system size on the studied transitions: for $m=1$
the results for the protocol (I) [protocol (II)] simulations are reported in Fig. \ref{rvsK_omega02}(b) 
for sizes ranging from $N=1,000$ to $N=10,000$. 
It is immediately evident that the transition from synchronized state to the asynchronous
state, following protocol (II), does not depend on N: for all considered sizes the transition 
happens in correspondence of $K \simeq 2$, analogously to what reported for
unimodal distributions \cite{olmi2014}.
Starting from the incoherent regime and following protocol (I) the system
reveals a jump to a finite $\bar r$ value for critical couplings increasing with $N$,
quite similar once more to the results reported for unimodal distributions.
We can conclude this sub-section by affirming that the phenomenology seen for
bimodal, but largely overlapping, distributions should not differ much from the one observed 
for unimodal distributions.

\section{Linear Stability Analysis}
\index{maximal Lyapunov exponent} \index{linear stability analysis}

To better characterize the synchronization transitions and the stability
of the observed states it is worth estimating the maximal Lyapunov exponent $\lambda_M$
following protocol (I) and (II) for an unimodal and a bimodal distributions. 
This quite time consuming analysis has been performed for 
a single inertia value $m=6$ and a single system size $N=1,000$, the scaling
of $\lambda_M$ with $N$ will be discussed in the following for specific coupling
constant values. 

In general, we observe that once the system fully synchronizes, $\lambda_M$
vanishes; therefore for most of the simulations associated to protocol (II)
corresponding to fully synchronized cluster down to the desynchronization transition,
$\lambda_M$ is zero. This is not the case for protocol (I) simulations
which reveal a positive $\lambda_M$ as soon as $\bar r$ is non zero.
Thus indicating that not only the dynamics characterized in terms of the
macroscopic order parameter $\bar r$ is hysteretic, but also at the level
of the microscopic dynamics, investigated via $\lambda_M$, the system has a
clear hysteretic behavior.

\begin{figure}
\centering
\includegraphics[angle=0,width=5.7cm]{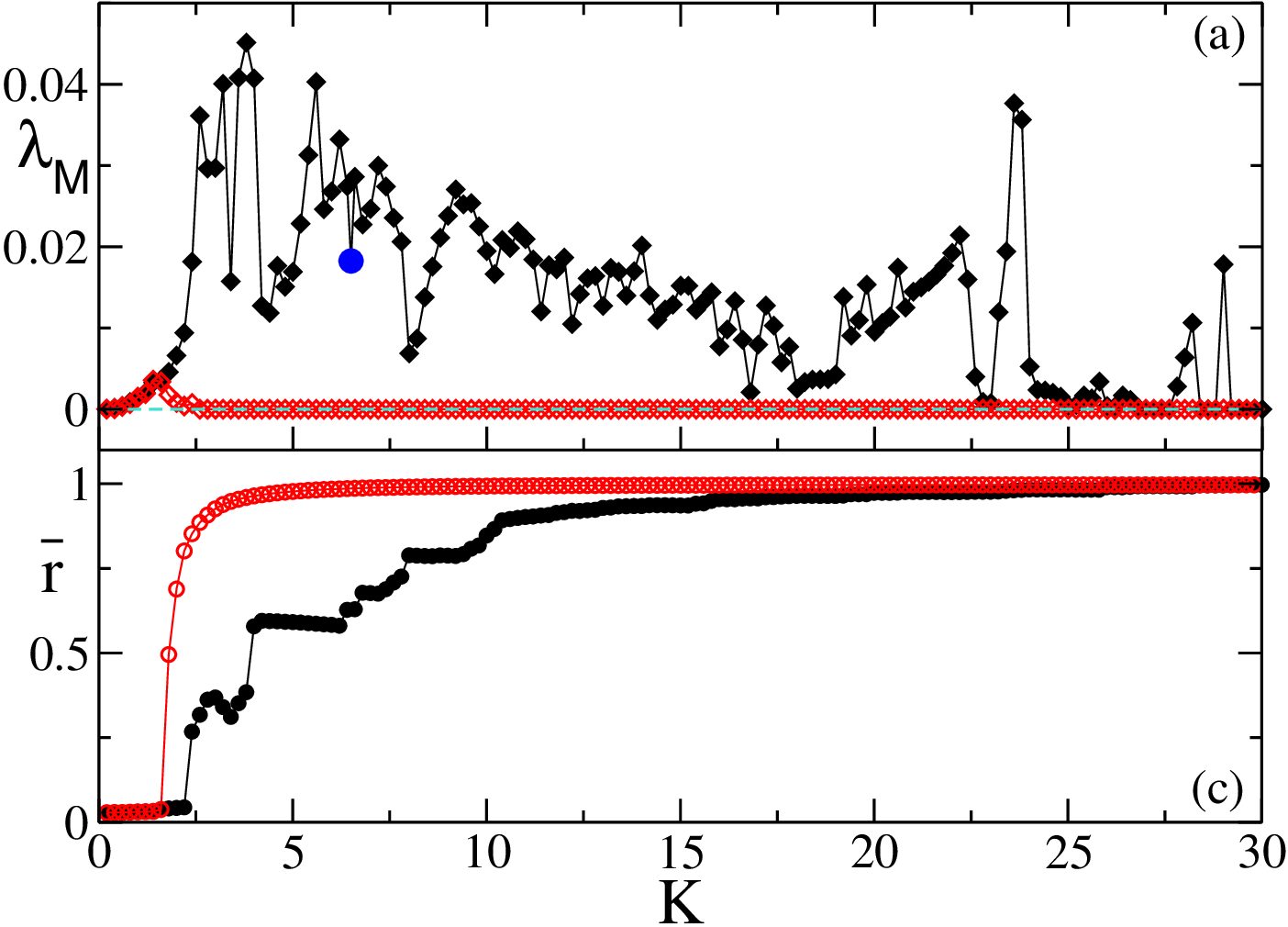}
\includegraphics[angle=0,width=5.7cm]{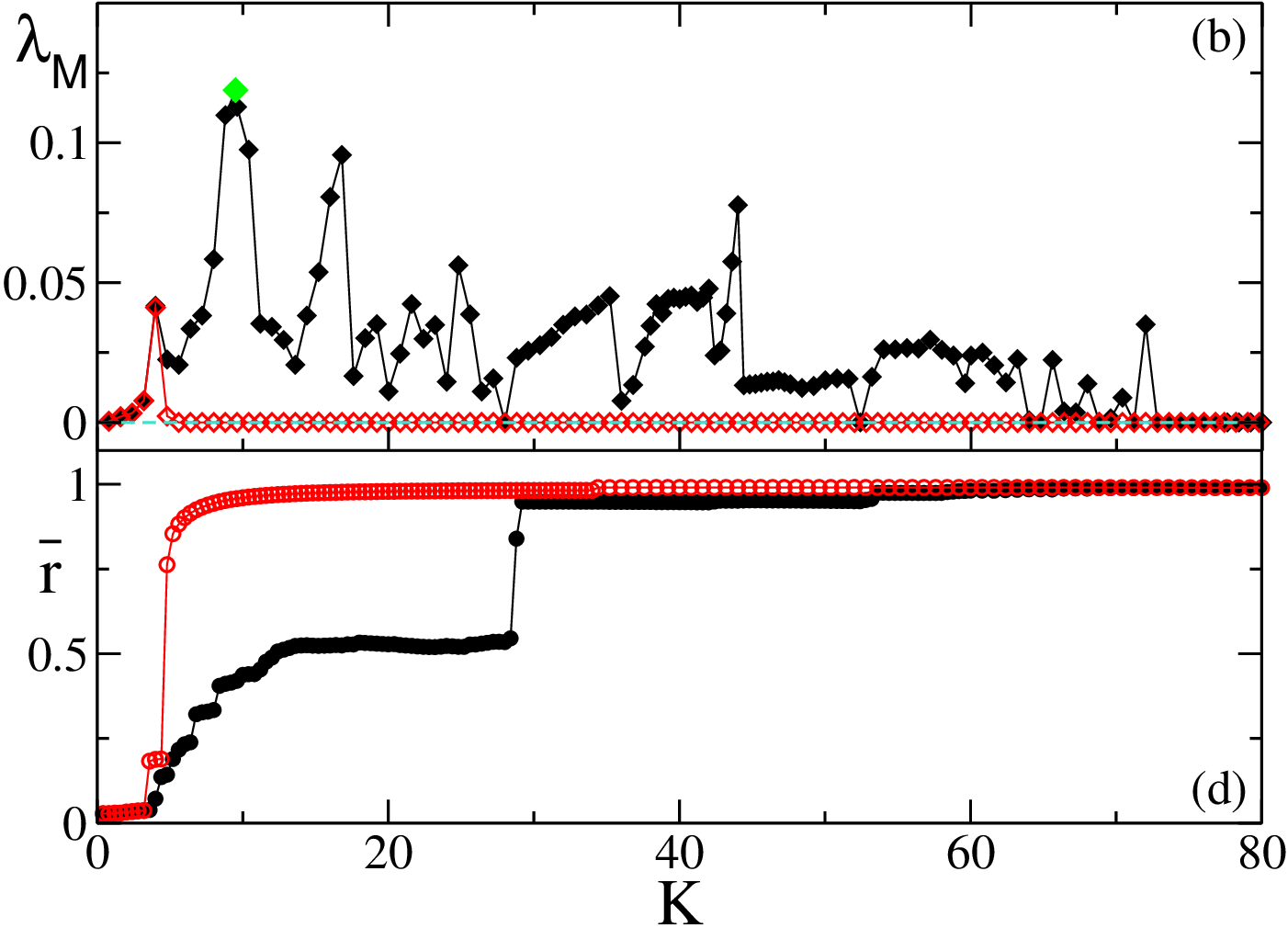} 
\includegraphics[angle=0,width=5.7cm]{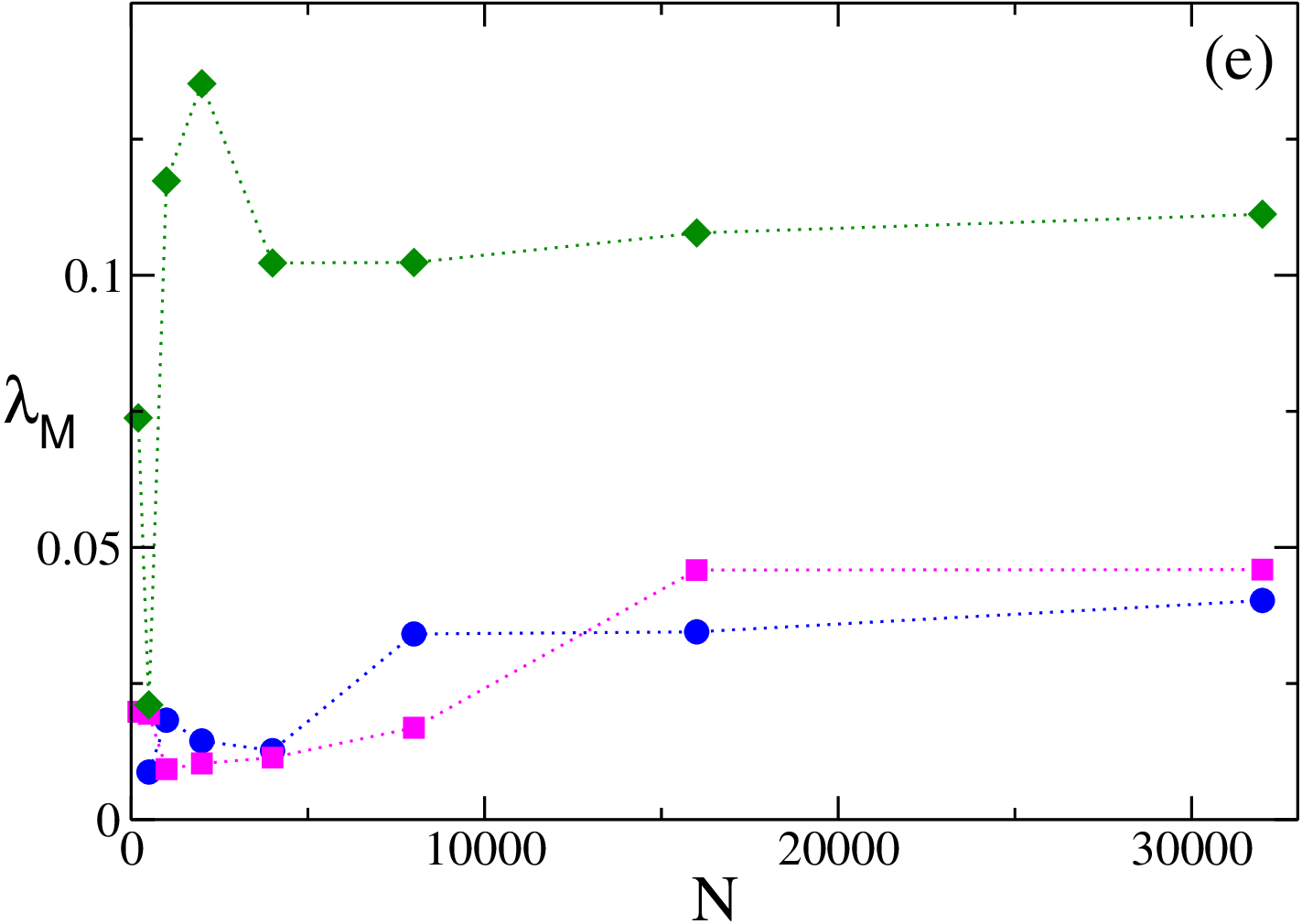} 
\caption{Maximal Lyapunov exponent $\lambda_M$ and the average order parameter ${\bar r}$
versus $K$ for unimodal (bimodal with $\Omega_0=2$) are shown in panel (a) 
(panel (b)) and in panel (c) (panel (d)), respectively.
The numerical data have been obtained by following protocol (I) (black circles) and then protocol 
(II) (red diamonds) from $K=0$ up to $K_M=30$ ($K_M=80$) with $\Delta K=0.2$ ($\Delta K=0.8$).
For panels (a),(c) $T_R=500$, and $T_W=50,000$; for panels (b),(d) $T_R=500$, and $T_W=400000$.
The different symbols in (a) and (b) denote the value for which the further analysis reported
in panel (e) has been done. 
Panel (e): $\lambda_M$ versus $N$ for different couplings and
frequency distributions. Blue circles  refer to unimodal distribution and coupling 
constant $K=6.5$; magenta squares (green diamonds) refer to binomial distributions with 
$\Omega_0=0.2$ and $K=6.7$ ($\Omega_0=2$ and $K=9.5$).
$\lambda_M$ has been averaged over a time window $T_W=4,000-400,000$, 
after discarding a transient time $T_R=1,000 - 10,000$ depending on the system size. 
For all panels $m=6$ and $N=1,000$.
\label{lyapmax}}
\end{figure}

The behavior of $\lambda_M$ with $K$ exhibits chaotic dynamical
states with windows of regularity for both unimodal and bimodal distribution with $\Omega=2$,
as shown in Figs.~\ref{lyapmax} (a) and (b).  As a general aspect, we observe 
the maximal level of chaoticity immediately after the transition from the incoherent state
to partially coherence, where small clusters of synchronized oscillators and
drifting oscillators coexist. The increase of $\bar r$ is accompanied by a
trend of $\lambda_M$ to decrease and finally to vanish for $\bar r \to 1$

An important aspect to understand is if this dynamics is {\it weakly} chaotic or not,
in particular this amounts to verify if, in the thermodynamic limit,  $\lambda_M$
will vanish or will remain finite. In order to test for this aspect, we have considered
a configuration obtained by following protocol (I) for a specific coupling
and analyzed $\lambda_M$ versus the system size for $ 200 \le N \le 32,000$.
The results for unimodal distributions, as well as for bimodal
ones with $\Omega_0 =2 $ and $\Omega_0 =0.2 $ are shown in Fig.~\ref{lyapmax} (e).
It is clear for all the considered cases that the system remains chaotic for
diverging system sizes.
 
\begin{figure}
\centering
\includegraphics[angle=0,width=5.7cm]{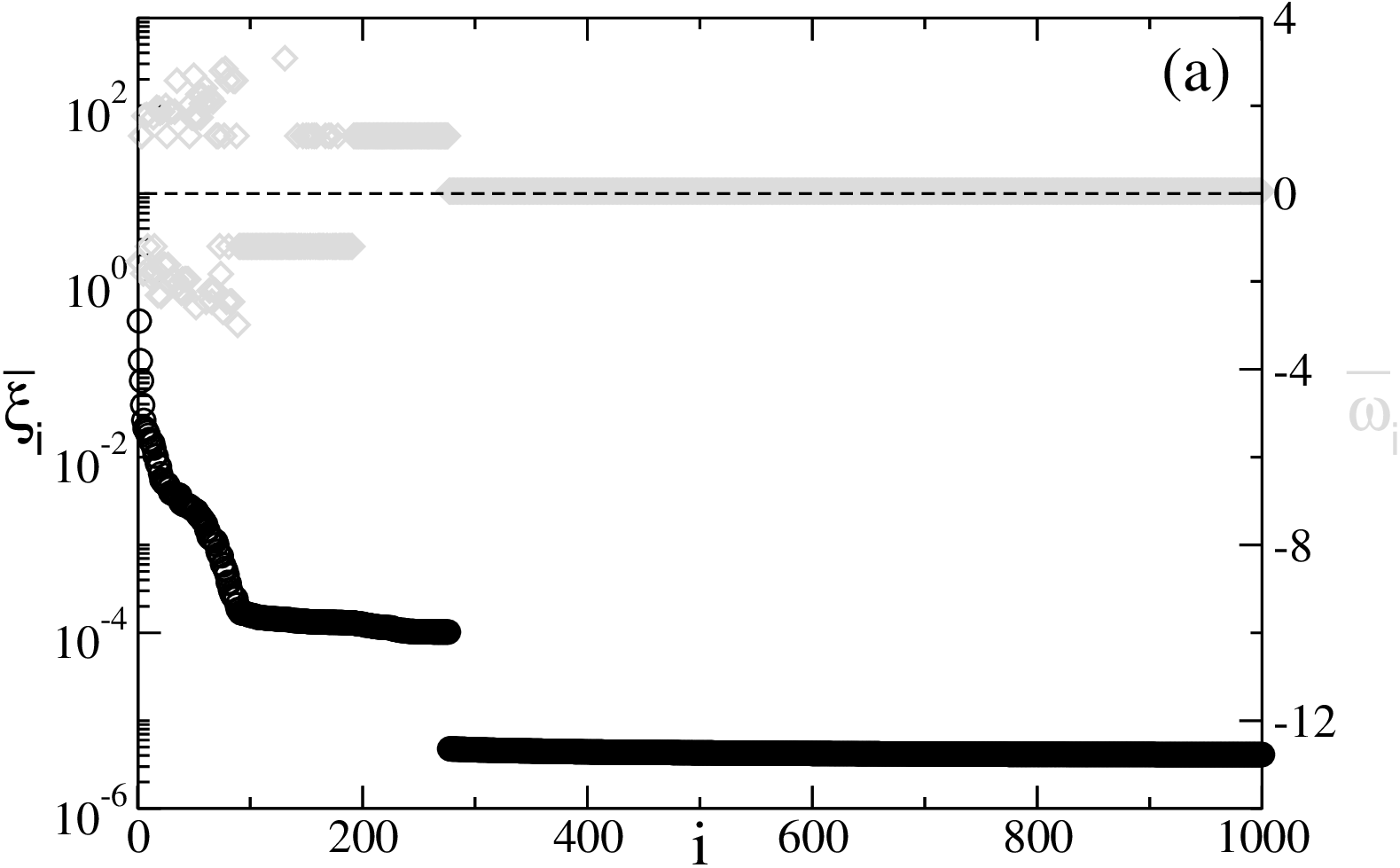}
\includegraphics[angle=0,width=5.7cm]{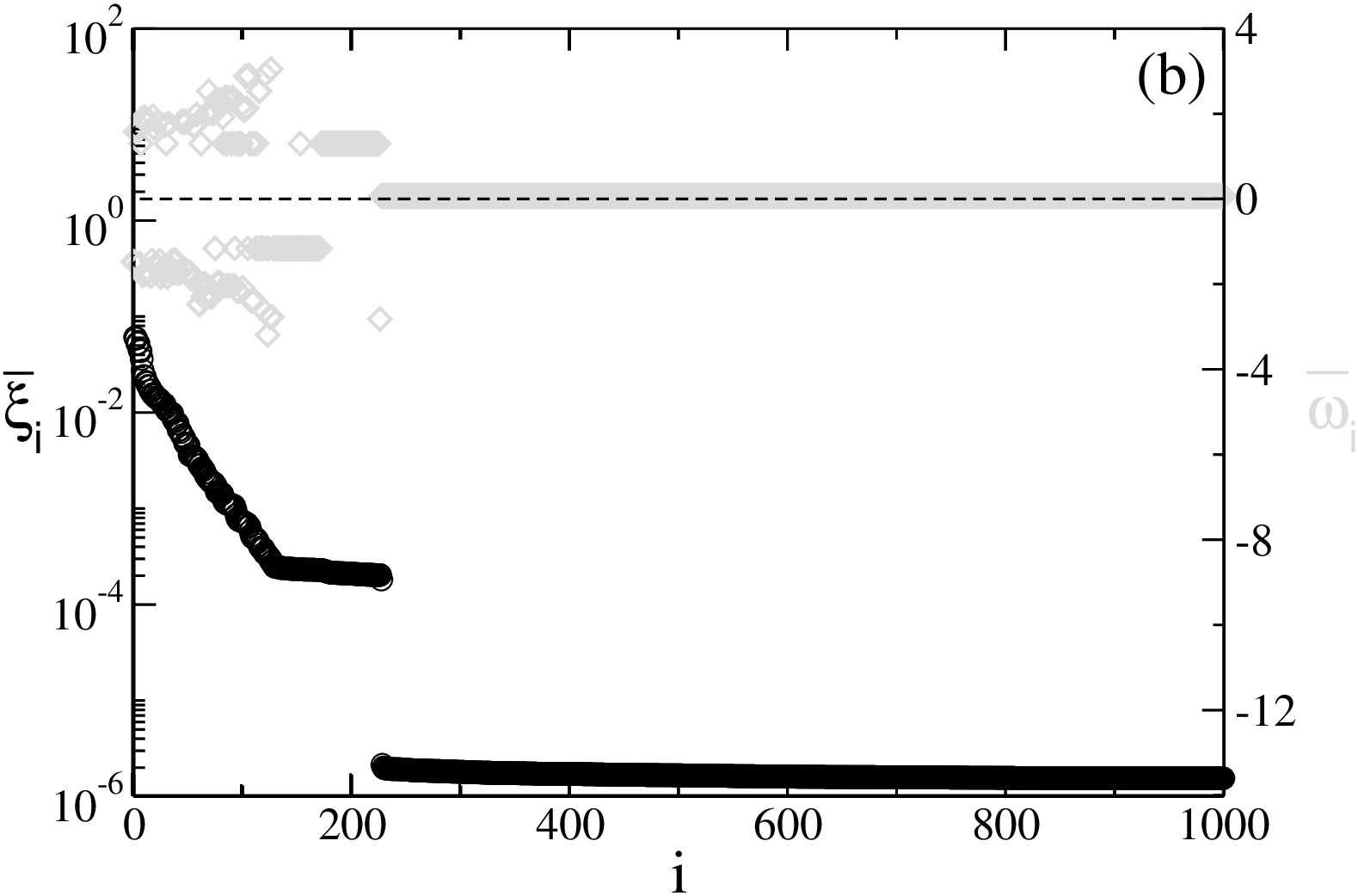} 
\includegraphics[angle=0,width=5.7cm]{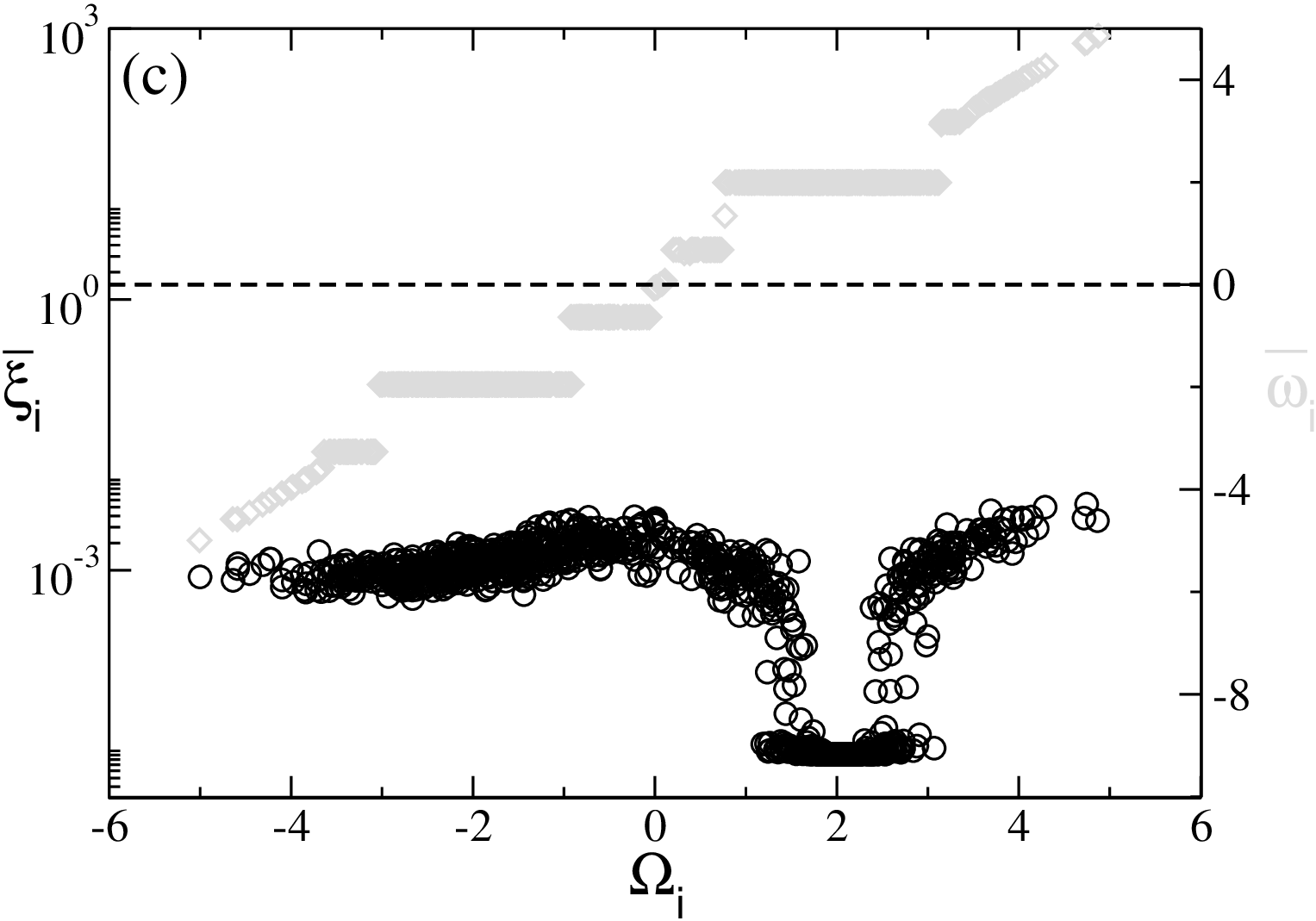}
\caption{Average squared amplitude of the components of the maximal Lyapunov vector 
$\bar \xi_i$ (black circles) and average frequencies $\bar \omega_i$ (grey diamonds)
versus the rotator index for unimodal (a) and bimodal distribution with $\Omega_0=0.2$ (b).
In both panels the oscillators are ordered according to the values of $\bar \xi_i$.
Panel (c): $\bar \xi_i$ (black circles) and average frequencies $\bar \omega_i$ (grey diamonds)
of the oscillators as a function of their natural frequency for bimodal distribution with 
$\Omega_0=2$. Panel (a) refers to coupling constant $K=6.5$, 
panel (b) to $K=6.7$ and panel (c) to $K=9.5$.
For all panels inertia $m=6$ and $N=1,000$, $T_R=500$, and $T_W=400000$.
\label{lyapvec}}
\end{figure}

As a final aspect we would like to understand which oscillators contribute more 
to the chaotic dynamics of the system; this can be understood
by measuring the average squared amplitude of the components of 
the maximal Lyapunov vector~\index{Lyapunov vector} $\bar \xi_i$ (see the definition 
reported in Eq.~\ref{loc_vect}).  In particular, we consider the three cases 
analyzed in Fig.~\ref{lyapmax} (e) for $N=1,000$. The  corresponding results are 
shown in Fig.~\ref{lyapvec}. From panel (a) and (b) of the figure it is clear that for
the unimodal distribution, as well as for the largely overlapping bimodal distributions,
the chaotic activity is associated almost exclusively to the rotators which are outside the large
clusters of locked oscillators with $\bar \omega_i \simeq 0$. Thus
confirming recent results reported for two coupled populations of
rotators with identical natural frequencies~\cite{olmi2015}. 

However, the situation for the bimodal distribution with $\Omega_0 = 2$ is
different; in particular, as shown in Fig.~\ref{lyapvec} (c), the network
for this large value of the inertia and the considered coupling does not
exhibit a cluster of locked oscillators with  $\bar \omega_i \simeq 0$, 
but only drifting clusters. In this case the
rotators outside and inside the clusters seem to contribute equally
to the maximal Lyapunov vector, with the possible exclusion of a group
of rotators with $\Omega_i \simeq \Omega_0$.

\section{Conclusions}

We have studied the synchronization transition for a globally coupled Kuramoto model with 
inertia for different frequency distributions. For the unimodal frequency distribution we 
have shown that clusters of locked oscillators of any size coexist within the hysteretic region. 
This region is delimited by two curves in the plane individuated by the coupling and the average 
value of the order parameter. Each curve corresponds
to the \textit{synchronization} (\textit{desynchronization}) profile obtained starting from the  
fully desynchronized (synchronized) state. For sufficiently large inertia values, clusters composed 
by drifting oscillators with opposite velocities (standing wave state) emerge in addition to the 
locked oscillators clusters. The presence of clusters of whirling rotators induces oscillatory 
behavior in the order parameter.

For bimodal frequency distribution the scenario can become more complex since it is possible to play 
with an extra parameter: the distance between the peaks of the distributions. For simplicity we have 
analyzed only two cases: largely overlapping distributions ($\Omega_0=0.2$), and almost not overlapping 
distributions ($\Omega_0=2$).
The phenomenology observed for $\Omega_0=0.2$ resembles strongly that found for the unimodal distribution. 
The analysis of the non overlapping case reveals new interesting features. In particular, the transition 
from incoherence to coherence occurs via several states: namely, traveling waves, standing waves and 
finally partial synchronization. 
This scenario resembles that reported for the usual Kuramoto model for a bimodal distribution
~\cite{crawford1994,martens2009,pazo2009}. However, in our case the transition is always largely hysteretic, 
and for non overlapping distributions, traveling waves are
clearly observable at variance, not only with the results for the Kuramoto model
~\cite{crawford1994,martens2009,pazo2009}, but also with the theoretical phase diagram reported 
in~\cite{acebron2000} for oscillators with inertia.
A peculiar aspect is that in the thermodynamic limit we expect a direct discontinuous
jump from the incoherent to the coherent phase, without passing through any intermediate state. 
The critical coupling $K^{PS}$ required to pass from incoherence to partial synchronization is 
independent of the system size and grows linearly with inertia, while the partially synchronized 
state looses its stability at a smaller coupling $K^{DS} < K^{PS}$ which is the same for any inertia 
value and  system size. 

Finally, by performing a linear stability analysis we have been able to show that the hysteretic 
behavior is not limited to macroscopic observables, as the level of synchronization, but it is 
revealed also by microscopic indicators as the maximal Lyapunov exponent. In particular, we expect 
that in a large interval of coupling values chaotic and
non chaotic states will coexist.

%


\section*{Acknowledgments}

We would  like to thank  E.A. Martens, D. Paz\'o, E. Montbri\'o, M. Wolfrum
for useful discussions. We acknowledge partial financial support from
the Italian Ministry of University and Research within the project CRISIS LAB PNR 2011-2013.
This work is part of of the activity of the Marie Curie Initial  Training Network 'NETT' project 
\# 289146 financed by the European Commission.

\bibliography{ref}

\printindex
\end{document}